\documentclass[]{article}

\usepackage{graphicx}
\usepackage{cancel}
\usepackage{authblk}
\usepackage{amsmath}
\usepackage{amsthm}
\usepackage{amssymb}

\usepackage[sort&compress,numbers]{natbib}
\usepackage[colorlinks=true,linkcolor=black, citecolor=blue, urlcolor=blue]{hyperref}

\begin{document}

\title{On the influence of powder particle size on single-track formation in  laser powder bed fusion of AlSi10Mg alloy} 

\author[1,2]{Y. Solyaev}
\author[2]{V. Dobryanskiy}
\author[3]{Nguyen Long}
\author[4]{Stanislav Chernyshikhin}

\affil[1]{Institute of Applied Mechanics of Russian Academy of Sciences, Moscow, Russia}
\affil[2]{Moscow Aviation Institute, Moscow, Russia}
\affil[3]{Le Quy Don University of Science and Technology, Hanoi, Vietnam}
\affil[4]{National University of Science and Technology MISIS, Moscow, Russia}

\setcounter{Maxaffil}{0}
\renewcommand\Affilfont{\itshape\small}

\date{\today}

\maketitle

\begin{abstract}
In this paper, we present the results of single-track experiments conducted for different fractions of standard AlSi10Mg powder, which were sieved to achieve varying mean size of the particles. We observed strong differences in the melting behaviour of fractions at relatively low levels of input energy in laser powder bed fusion (LPBF) process. Namely, the remarkable particle size effect arises for the position of lack of fusion boundary, i.e. for the range of process parameters where the laser power becomes sufficient for complete through-thickness melting of the powder layer at given laser scanning speed. We established that this boundary corresponds to an approximately constant linear energy density at low levels of Peclet number (Pe $<$ 2), while the constant enthalpy density defines this boundary at the higher levels (Pe $>$ 2). Specifically, when the mean particle size ranges from 28 to 64 $\mu$m, the required linear energy density for stable track formation ranges from 50 to 167 J/m and the nominal enthalpy density ranges from 4.4 to 15 J/mm$^3$.
Based on the scaling law analysis and numerical simulations, we show that observed phenomena can be attributed to the change of absorptivity of powder layer, which depends on particle size and packing density. Also, we show that the values of absorptivity identified based on the analysis of position of lack of fusion boundary (0.13-0.38 for powder size 64-28 $\mu$m) correlate well with those found from the analysis of melt pool width in the formed single-tracks. 

\end{abstract}

\section{Introduction}

The influence of powder granulometry on the feedstock and part performance in laser powder bed fusion (LPBF) processes has been extensively studied \cite{spierings2011influence, tan2017overview}. It was shown that the wider particle size distribution (PSD) results in higher packing density and corresponding increase in the effective thermal conductivity of the powder \cite{rombouts2005photopyroelectric,yadroitsev2010single}, decrease of the effective thickness of powder layers \cite{spierings2009} and increase of parts density and ductility \cite{tan2017overview,riener2020influence}. The use of fine powders as the main feedstock (or as the additions to standard feedstock) with appropriate choice of optimal LPBF parameters allowed to increase the part density \cite{spurek2021influence,spurek2024effect}, as well as mechanical strength \cite{spierings2011influence}, ductility \cite{chu2024improved}, and surface quality \cite{spierings2011influence, riener2020influence} of fabricated metal samples. The use of fine powders  also allowed to improve LPBF capabilities for the manufacturing of thin-walled structures \cite{calignano2018manufacturing}, though it can be limited by lower flowability of fine powders and high tendency to agglomeration of very small particles \cite{balbaa2021role,ziri2022combined,suvorova2025laser}. 

The improvement of melting performance of powders with smaller mean size of the particles is usually observed in the experiments and attributed to their higher absorptivity, since the fine powders exhibit larger surface area as compared to coarse powders \cite{simchi2004role,spierings2011influence, tran2017analysis, riener2020influence}. 
From theoretical point of view, the decrease of mean size of powder particles results in the higher optical thickness of the powder layer with corresponding increase of laser extinction coefficient and multiple scattering effect inside the layer, and more uniform irradiation of the layer surface \cite{Gusarov2009, tan2017overview, yang2018laser, zhang2019influence}. 
Strong particle size effects on the melting performance of powder bed can be also predicted theoretically based on the phenomenological high-grade models of heat transfer in granular medium \cite{solyaev2022gradient}.
Thus, varying the mean size of particles and PSD one can achieve the change of powder absorptivity. 
For example, for tungsten powder it was shown that the  absorptivity increases from 0.5 to 0.6 with decrease of particle size from 45 $\mu$m to 5 $\mu$m \cite{zhang2019influence}. 
For maraging steel and aluminum powders, it was shown that for the mean size of particles equals to 25 $\mu$m and 60 $\mu$m, the absorptivity reduces from 0.78 to 0.65 (steel), and from 0.4 to 0.27 (aluminum), respectively. In Ref. \cite{yang2018laser}, it was shown that absorptivity of monomodal AlSi12 powders varies in the range 0.22-0.12 when the size of particles varies in the range 10-60 $\mu$m. 
In Ref. \cite{ziri2022combined} authors used hemispherical reflectance spectroscopy for 316L stainless steel powders, and it was shown that absorptivity increased in the range 0.7-0.79, when the mean size of particles ranges from 67 $\mu$m to 6 $\mu$m. 
Spectroscopy methods was also used in Refs. \cite{riener2020influence, balbaa2021role} to evaluate the absorptivity of AlSi10Mg powders of different size. It was found that absorptivity changes from 0.59 to 0.63 for the powders with $d_{50}$ equals to 31 $\mu$m and 41 $\mu$m, respectively \cite{riener2020influence}. Also, involving Kubelka-Munk theory, it was shown that absorption of AlSi10Mg powder with $d_{50} = 9$ $\mu$m  is higher in 13\% than those one of the powder with $d_{50} = $ 40 $\mu$m \cite{balbaa2021role}.

Notable that the most of applied studies with AlSi10Mg powders are usually performed in more or less standard ranges of LPBF process parameters at rather high level of energy input (with laser power higher than $\sim$200 W) \cite{balbaa2021role,chu2023influence}. In these high-energy regimes, the powder layer melts fast and the laser irradiation interacts primarily with the melt pool surface, not with the powder particles \cite{ye2019energy, doring2025trans}. As a result, the minor effects of the particle size on the melting performance of AlSi10Mg powders and on the process efficiency were observed previously (see, e.g. \cite{chu2023influence}).

In the present study, we show that the strong change of absorptivity of fine AlSi10Mg powders can be directly observed in LPBF processes considering regimes with relatively low input energy, at which the melting is relatively slow and laser interactions with powder particles plays significant role. 
We show that there arise the remarkable shift of the lack of fusion boundary on the process map obtained by using single-track experiments conducted for the powders with different mean size of particles. 
Note that this particle size effect in single-track experiments with AlSi10Mg powders was not widely discussed previously. 
At the same time, such simple experiments allowed us to evaluate different melting behaviour of the powder fractions and to assess their absorptivity. Considering the position of lack of fusion boundary, we performed inverse analysis and evaluated the absorptivity of AlSi10Mg powder fractions with different mean size of particles. In the calculations, we used the non-dimensional scaling law analysis introducing the coefficient of single-track melting efficiency following Ref. \cite{wang2019dimensionless}. 
For relatively high levels of Peclet number (Pe) we additionally involve the criterion formulated with respect to the normalized enthalpy as suggested in Refs. \cite{hann2011simple, ye2019energy}.
Also, we provided numerical simulations with direct definition of laser extinction in the powder layer by using an approach that was developed in Ref. \cite{Gusarov2009} and used recently in Ref. \cite{yang2022validated}. 
Moreover, we found that the values of absorptivity identified based on the supposed inverse analysis for the powder melting performance correlate well with those found from the analysis of the melt pool width in the formed single-tracks.
The last one is conducted by using the methodology developed recently in Ref. \cite{doring2025trans}. All considered approaches show that in the case of relatively low levels of energy input, we can achieve the strong influence of particle size effect on single-track formation and possibly exploit it to increase the effective absorptivity of powder feedstock in LPBF processes.

Note that in the present study we use the term "lack of fusion" boundary within the analysis of single-track experiments, meaning that the energy input  becomes sufficient to provide the through-thickness melting of powder layer and form the single-track that is well fused to the substrate. This boundary on the process map can be also denoted as "under-melted threshold" \cite{agrawal2022predictive} or the boundary of the "continuous track formation" \cite{wang2019dimensionless}. Although these alternative terms would be not exactly correct in the present case, since the "under-melted threshold" (the beginning of melting) assumes the lower energy input  and the "continuous track formation" (the track is fused to the substrate and no balling effects occurs) assumes the higher energy input as compared to the present analysis. The position of these boundaries on the obtained process maps demonstrates strong dependence on the powder size.

Note that similar effects were established in Ref. \cite{ziri2022combined} in the experiments with 316L stainless steel powders, where the influence of powder particle size on the lack of fusion and keyhole boundaries was found. However, no strict monotonic trend was observed in the relationship between the required level of energy input and particle size for achieving high-quality single-tracks. Specifically, both finer and coarser 316L powders required higher energy input compared to the standard powder \cite{ziri2022combined}. In the present study, for AlSi10Mg powders, we observed a minor influence of particle size on the keyhole boundary and provided a theoretical explanation for the significant shifts of lack of fusion boundary. We show that the lack of fusion boundary  corresponds to approximately constant level of linear energy density (for Pe $<$ 2) or enthalpy density (for Pe $>$ 2) that monotonically increase for coarser powders.

We can also refer to the findings of Refs. \cite{ye2019energy, wang2019dimensionless, yang2022validated, bogdanova2024mesoscale}, which demonstrate significant variations in powder absorption efficiency with changes in the laser beam diameter and the powder layer thickness.
It is worth noting that the nature of these size effects is similar to those one observed in the present study, since from the point of view of non-dimensional analysis the melting performance of powder bed does not depend on the absolute values of dimensional parameters but on their ratios. Thus, considering ratio between the thickness of powder layer $h$ and the mean diameter of the particles $d$ we can state that the higher optical thickness and absorptivity of powder layer will arise for higher values $h/d$ \cite{Gusarov2009}. Also, the size effect can be related to the change of ratio between the laser beam radius $R$ and the particles size $d$, so that more uniform irradiance of the powder bed will arise for higher $R/d$. Specifically, in Ref. \cite{ye2019energy} authors used Ti64 powder with mean size of particles $d\approx$ 30 $\mu$m, while the layer thickness was varied in the range $h=$ 50-100 $\mu$m. 
It was demonstrated that the measured absorptivity increases from 0.35 to 0.5 at low energy input as the ratio $h/d$ varies from 1.67 to 3.3. The maximum absorptivity at low laser power (0.5) almost approached the typical values for keyhole regimes (0.65), where no particle size effect was observed.
In the present study, we provide evidence for similar effects in the case of AlSi10Mg alloy, when the ratio $h/d$ ranges from 1.1 to 3. However, in our case the variation in this ratio is attributed to differences in the mean particle size ($d$) of the used powders.

 Practical application of presented results can be related to the development of LPBF processes with highly reflective and highly conductive materials, for which the search for effective methods to increase absorptivity is a relevant problem up to date \cite{sefene2022state}. As it follows from the presented analysis, for such materials, it can be advisable to use the finer fractions of powders and processing modes with lower energy input (with appropriate adjustment of other process parameters), where the particle size effects on absorptivity becomes significant. 
 Additionally, presented experimental results can be used for validation of advanced numerical particle-scale models of the melting behaviour of powder beds.

\section{Materials and methods}

\subsection{AlSi10Mg powders}

For experimental studies, we used one batch of commercially available gas-atomized AlSi10Mg powder (ASP-45, UC RUSAL). The powder was sieved according to ASTM B214 with the use of vibratory sieve shaker Analysette 3 (Fritsch, Germany) equipped with the sieves with different mesh width equals to 20 $\mu$m, 32 $\mu$m, 40 $\mu$m, and 50 $\mu$m (ISO 3310). Five fractions of the powder with different mean size of particles were obtained. Powder fraction P1 passed through the sieve with mesh width 20 $\mu$m. Fractions P2, P3, and P4 were collected between the sieves with mesh widths 20/32 $\mu$m,  32/40 $\mu$m, and  40/50 $\mu$m, respectively. Powder fraction P5 remained on the top of sieve with mesh width 50 $\mu$m. Initial non-sieved (standard) powder was also used in the experiments and it is denoted as P0. 

 \begin{figure}[b!]
 \centering
  \includegraphics[width=\linewidth]{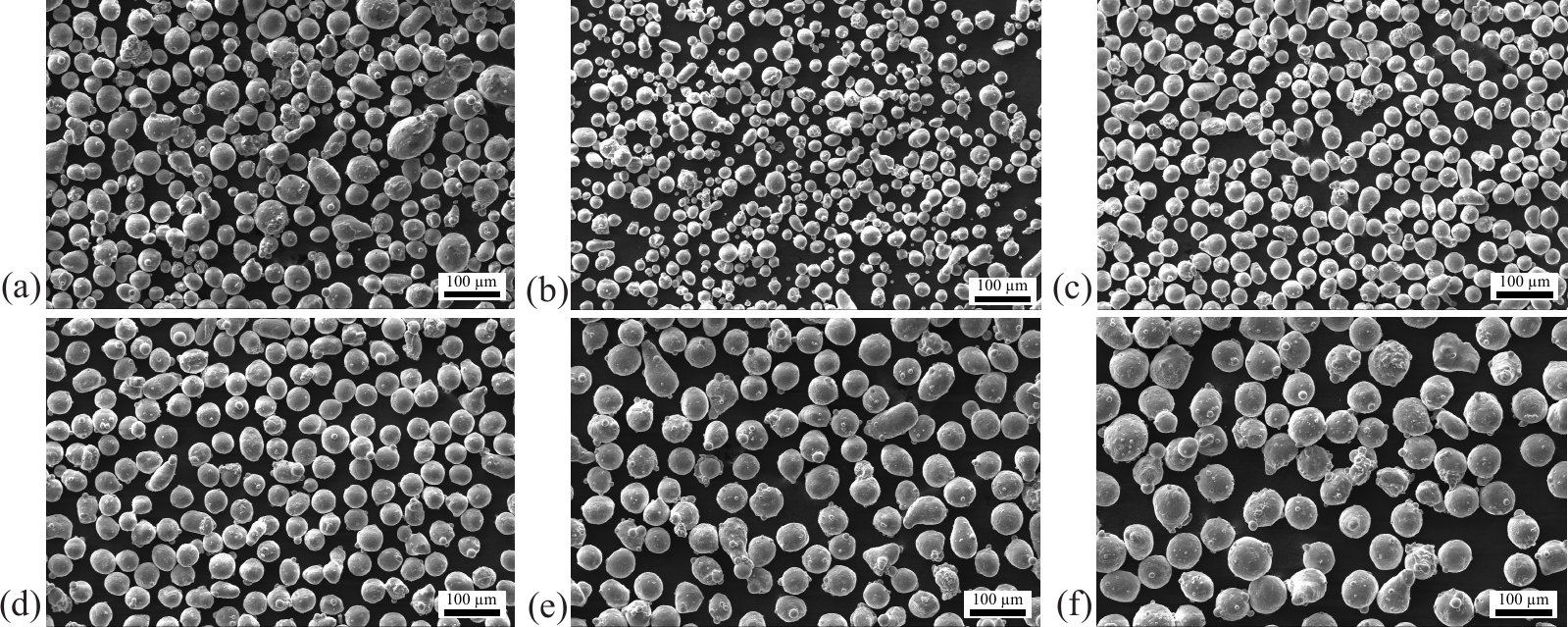}
  \caption{SEM images of standard AlSi10Mg powder P0 (a) and its sieved fractions P1 (b), P2 (c), P3 (d), P4 (e), P5 (f) \label{sem}}
\end{figure}

The powders granulometry and chemical composition were investigated by using scanning electron microscope EVO-40 (Carl-Zeiss, Germany) with micro-XRD analysis system X-Max (Oxford Instruments).  Obtained SEM images of P0-P5 powders are presented in Fig. \ref{sem}. Most of powder particles have a shape close to spherical form. For coarser powders, some small satellites and oxide flakes attached to the surface of particles can be observed (Fig. \ref{sem} e,f). 
It was established that chemical composition of the used standard powder (P0) is in accordance with DIN EN 1706, though some variation of Si content was found for the powder fractions, so that coarser powders (larger particles) contains higher amount of silicon. 
Similar observations for AlSi10Mg powder fractions has been discussed in Refs. \cite{balbaa2021role,dobryanskiyinfluence}. The change of Si amount in powder particles can be attributed to surface effect or to the higher probability for presence of Si precipitates in the large size particles. Established  variation of chemical composition does not strongly affect the physical properties of the powders. Namely, by using differential scanning calorimetry performed by using NETZSCH STA 449 F1 (Jupiter, Germany) it was found that solidus/liquidus temperature equals to 563/597 $^o$C with variation less than 2$^o$C for standard (P0) and for all sieved powders (P1-P5).

The particle size distribution (PSD) was evaluated by using obtained SEM images according to ISO 13322. Apparent density $\rho_a$ of the powders were measured according to ISO 3923 by using flowmeter HFLOW-1 (Bettersize Ltd., China). The packing density of the powders were calculated as $f=\rho_a/\rho$, where $\rho=2670$ kg/m$^3$ is the density of solid AlSi10Mg alloy. 
Notable that the packing density of small-thickness powder layers usually differs from those of volumetric samples of the powder \cite{spurek2024effect,riener2020influence}.
Therefore, all measurements were performed for the initially sieved powder samples as well as for additionally collected samples obtained by using special equipment for simulation of thin-layer packing of the powder fractions in LPBF process \cite{riener2020influence}.
In order to conduct such experiments, we used small aluminum platform with a cavity, which depth was 120 $\mu$m and diameter was 30 mm. The depth of the cavity was chosen to be enough small to asses the thin-layer density effects and at the same enough large to fabricate it with high dimensional accuracy with available equipment. The powder were coated over the cavity and the masses of platform with ($M_1$) and without ($M_0$) the powder layer were measured by using laboratory balance Entris II (Sartorius, Germany) with accuracy $ 10^{-4}$ g. Apparent density of powder layer was estimated by using the found mass of the layer ($M_1-M_0$) and known volume of the cavity. The powder layer packing density was then calculated in a similar way as described above for the volumetric packing density. Ten repeated tests were performed for layer density analysis for each powder P0-P5. 

\begin{figure}[t!]
 \centering
  (a)\includegraphics[width=0.43\linewidth]{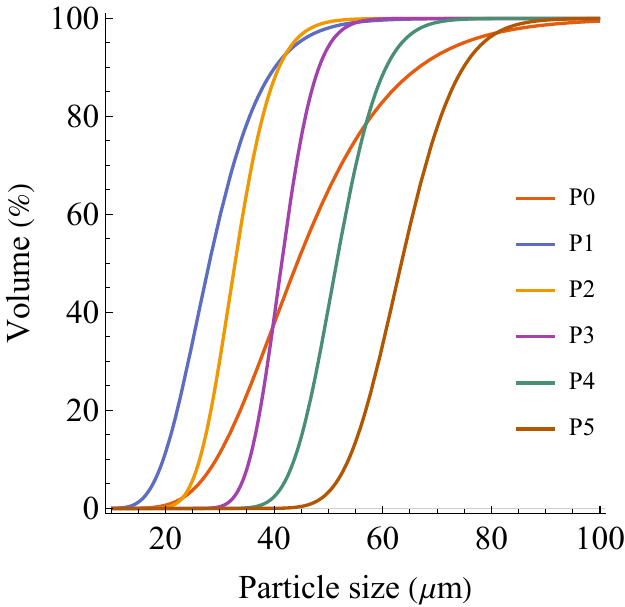}\quad
  (b)\includegraphics[width=0.43\linewidth]{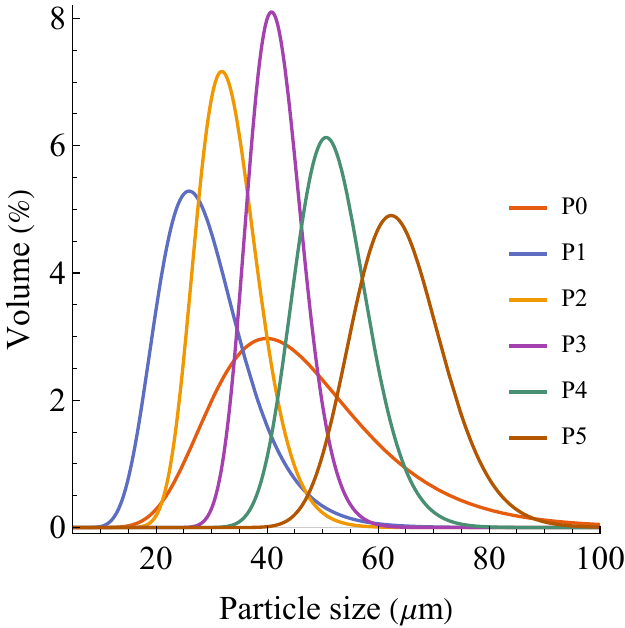}
  \caption{Particle size distribution for standard AlSi10Mg powder (P0) and its sieved fractions (P1-P5) with different mean size of particles, (a): cumulative distribution, (b): frequency distribution. }
  \label{psd}
\end{figure}

 Established granulometric parameters of standard (P0) and sieved (P1-P5) powders are presented in Table \ref{gran}. The mean particle size of the powders (assessed via $d_{50}$) varies in the range 28-64 $\mu$m. Notable that $d_{50}$ values do not strongly change between the volume and layer samples of the powder. At the same time the packing density of the powders becomes lower in the  powder samples obtained from thin-layer analysis. Moreover, there arise some dependence of packing density on the mean size of particles: the sieved finer fractions have lower packing density and the layer density of standard powder is in between those one of coarser and finer fractions. Note that similar dependences were established in Ref. \cite{riener2020influence}, though the inverse dependences have been also reported (see, e.g. \cite{spurek2024effect}) that can be determined by the PSD peculiarities of different sieved fractions. Measured PSD are shown in Fig. \ref{psd}. The data in this figure is presented for powder samples collected by using equipment for the thin-layer packing density analysis. Log-normal approximations were used for the experimentally measured distributions of particle diameters. It can be seen that the standard powder (P0) has wider PSD in comparison with its fractions (P1-P5).

\begin{table}[t!]
\centering
\caption{Granulometric parameters measured for both volume and layer samples of AlSi10Mg powders P0 (standard) and P1-P5 (sieved fractions)}
\scriptsize
\begin{tabular}{ccccccccccc}
\hline
Powder & \multicolumn{2}{c}{$d_{10}$ ($\mu$m)} & \multicolumn{2}{c}{$d_{50}$ ($\mu$m)} & \multicolumn{2}{c}{$d_{90}$ ($\mu$m)} & \multicolumn{2}{c}{Span} & \multicolumn{2}{c}{Packing density, $f$} \\
type & Volume & Layer & Volume & Layer & Volume & Layer & Volume & Layer & Volume & Layer \\
\hline
P0 & 33.6 & 29.4 & 48.4 & 44.2 & 69.7 & 66.6 & 0.75 & 0.84 & 0.542& 0.432\\
\hline
P1 & 17.2 & 19.6 & 25.8 & 28.1 & 38.6 & 40.2 & 0.83 & 0.73 & 0.530& 0.380\\
\hline
P2 & 27.1 & 26.3 & 32.4 & 32.8 & 38.8 & 40.9 & 0.36 & 0.45 & 0.528& 0.421\\
\hline
P3 & 35.7 & 35.5 & 41.8 & 41.4 & 48.8 & 48.3 & 0.31 & 0.31 & 0.533& 0.433\\
\hline
P4 & 42.5 & 43.7 & 53.7 & 51.5 & 67.9 & 60.6 & 0.47 & 0.33 & 0.539& 0.473\\
\hline
P5 & 53.9 & 53.7 & 65.9 & 63.4 & 80.5 & 74.9 & 0.40 & 0.33 & 0.545& 0.508\\
\hline
\end{tabular}
\label{gran}
\end{table}

\subsection{Single-track experiments}

Single-track experiments were performed by using selective laser melting process with AddSol D50 system (Additive Solutions, Russia) with 400 W ytterbium-fiber laser (IPG Photonics) with wavelength 1064 nm and laser spot size $2R=$ 80 $\mu$m. Single tracks were formed on aluminum substrate in an argon atmosphere. The thickness of powder layer was chosen to be $h=$60 $\mu$m according to an assessment for the effective layer thickness in standard LPBF process \cite{ziri2022combined,spierings2009comparison}, i.e. $h=h_0/f$, where $h_0=$ 30 $\mu$m is the nominal layer thickness and $f$ is powder packing density that is approximately equals to 0.5 (see Table \ref{psd}).

Single tracks were fabricated within a wide process window with laser power $P$ = 40-400 W (with step 36 W) and scanning speed $v$ = 100-1980 mm/s (with step 235 mm/s). In total, 99 printing parameter regimes were investigated across all powder types (P0-P5). Single tracks morphology on the top view images was investigated by using optical microscope Saike Digital SK2700HDMI-AF (Saike, China). Then the substrates were cut at the centres of single-tracks, polished and etched. Evaluation of melt poll dimensions on the micro-section images were performed by using confocal laser microscope Olympus LEXT OLS5000 (OLYMPUS, Japan).

\subsection{Modelling methods}

\subsubsection{Scaling law analysis}

For the first assessments on the powders abosroptivity, we used the dimensionless analysis supposed in Ref. \cite{wang2019dimensionless} for evaluation of the energy input required for stable track formation. The following definition for the coefficient of single-track melting efficiency was used:
\begin{equation}
\label{n}
	N = \frac{E_v}{H_s}  
	=\frac{AP}{\pi\rho c(T_{m}^{eff}-T_0)h v\omega}
\end{equation}
where $E_v = A P \tau /(\pi \omega^2 h)$ is the volumetric density of input energy, $H_s = \rho c(T_{m}^{eff}-T_0)$ is the energy (enthalpy) density required for melting, $A$ is absorptivity, $P$ is laser power, $\omega=\sqrt{2}R$ is an effective radius of Gaussian laser beam, $\tau = \omega/v$ is the characteristic time, $v$ is laser scanning speed, $h$ is powder layer thickness, $T_m^{eff} = T_l + L_m/c$ is the effective melting point defined via the liquidus temperature $T_l$, the latent heat of melting $L_m$ and the specific heat capacity $c$; $T_0$ is the initial temperature of the powder layer and substrate.

The criterion for continuous track formation is defined by relation $N>1$ meaning that the input energy is enough to provide melting of the irradiated powder volume with radius $\omega$ and thickness $h$. 
The lack of fusion boundary can be defined on the process map with various laser power/speed as the curve where $N=1$ \cite{wang2019dimensionless}. If the experimental lack of fusion boundary can be approximated by constant linear energy density, then we can use relation \eqref{n} to evaluate the absorptivity as follows:
\begin{equation}
\label{a}
	A = \frac{\pi \rho c(T_{m}^{eff}-T_0)h\omega v^*}{P^*} 
	= \frac{\pi \rho c(T_{m}^{eff}-T_0)h\omega}{E^*_l}
\end{equation}
where $E^*_l = P^*/v^*$ is the critical linear energy density evaluated based on the established critical values of laser power $P^*$ and scanning speed $v^*$ that define the position of lack of fusion boundary on the experimental process map.

Based on the performed analysis (see next Section) we found that the lack of fusion boundary in the experiments with AlSi10Mg powders can be approximated based on the energy-based criterion \eqref{n} only at relatively low levels of laser scanning speed $v$.  For the higher laser speed, we found that it would be better to considered another criterion formulated with respect to the normalized enthalpy density \cite{agrawal2022predictive,doring2025trans,ye2019energy} that is given by:
\begin{equation}
\label{b}
	\beta = \frac{\Delta H}{H_s}  
	=\frac{AP}{\pi \rho c(T_{m}^{eff}-T_0)\sqrt{\alpha v \omega^3}}
\end{equation}
where $\Delta H = AP/(\pi\sqrt{\alpha v \omega^3})$ is the change of enthalpy due to heat input by laser power and $\alpha$ is material thermal diffusivity.

Note that the normalized enthalpy \eqref{b} defines the increase of surface temperature due to the action of moving heat source. Namely, the melting temperature will be reached when $\beta = 1$ \cite{agrawal2022predictive}. Therefore, assuming that this temperature-based criterion defines the position of lack of fusion boundary in the single track experiments at high scanning speed regimes, we can use \eqref{b} to evaluate the absorptivity of powder-substrate system as follows:
\begin{equation}
\label{ab}
	A = \frac{\pi \rho c(T_{m}^{eff}-T_0)\sqrt{\alpha v^* \omega^3}}{A\,P^*}
	 = \frac{\rho c(T_{m}^{eff}-T_0)}{\Delta H^*}
\end{equation}
where $\Delta H^*$ is the critical enthalpy density that corresponds to the position of lack of fusion boundary on the experimental process map.

In contrast to linear relation \eqref{a}, where $A \sim v^*/P^*$, in the obtained relation \eqref{ab} absorptivity depends on the critical process parameters as $\sim \sqrt{v^*}/P^*$. Therefore, the dimensionless criterion for the normalized enthalpy \eqref{b} defines the curvilinear shape of the lack of fusion boundary. 

The relations \eqref{a}, \eqref{ab} are used in the next section to evaluate the absorptivity of powders of different size, based on the established position of the lack of fusion boundary, which is determined by the sets of ($v^*$,$P^*$) values. 
Notable that similar distinct energy-based and temperature-based criteria have been discussed in Ref. \cite{oliveira2020revisiting} in view of the choice of optimal hatch distance in LPBF processes. 
In the present study, the approximation of the experimental lack of fusion boundaries was performed by using the least squares method under criteria $N=1$ \eqref{n} and $\beta=1$ \eqref{b} together with the corresponding relations for evaluation of absorptivity \eqref{a} and \eqref{ab}. These calculations are performed by using properties of liquid alloy (see Table \ref{tab1}) so that absorptivity $A$ remain single unknown parameter that is used to fit the theoretically predicted boundary to the experimental one. The value of $A$ that provides the best fit is considered as an assessment on the absorptivity of the powder of given size. It will be shown that criteria \eqref{n}, \eqref{b} are advisable to use for different scanning speed ranges. Specifically, to distinguish between low- and high-speed regimes, we will use the dimensionless Peclet number, defined as $Pe = 2\omega v/\alpha$ \cite{doring2025trans}.

\begin{table}[h!]
\caption{AlSi10Mg properties used in the simulations}
\small
\label{tab1}
\centering
  \begin{tabular}{lcc}
  \hline
    Parameters & Dimensions & Values \\
    \hline
    Density of solid, $\rho_s$ & kg/m$^3$ & 2670\\
    Density of liquid, $\rho_l$ & kg/m$^3$ & 1800\\
    Specific heat capacity of solid, $c_s$ & J/(kg K) & 922\\
    Specific heat capacity of liquid, $c_l$ & J/(kg K) & 1800\\
    Thermal conductivity of solid, $k_s$ & W/(m K) & 146\\
    Thermal conductivity of liquid, $k_l$ & W/(m K) & 120\\
    Latent heat of melting, $L_m$ & kJ/kg & 396\\
    Liquidus temperature, $T_l$ & $^o$C& 597\\
    Solidus temperature, $T_s$ & $^o$C& 563\\ 
    \hline
  \end{tabular}
\end{table}

\subsubsection{Numerical simulations}

\begin{figure}[t!]
 \centering
  \includegraphics[width=0.5\linewidth]{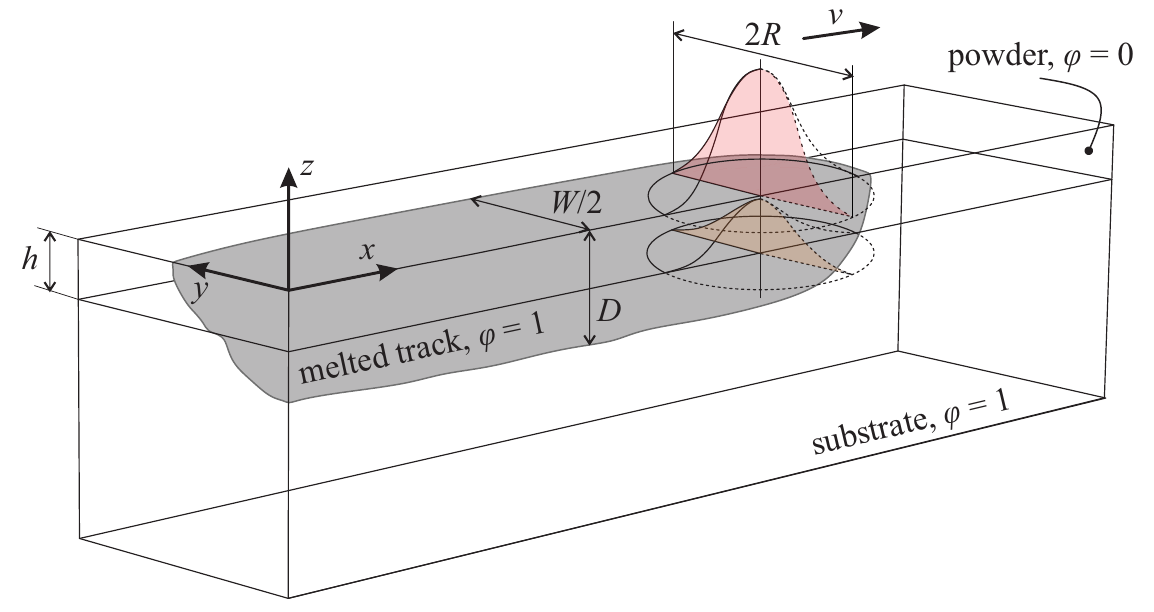}
  \caption{The illustration for the continuum model of powder layer/substrate system considered in the numerical simulations of the single track experiments}
  \label{fem1}
\end{figure}

For the refined analysis of a single-track formation, we also performed the numerical finite-element simulations based on the approach developed in Ref. \cite{Gusarov2009}. This method has been used recently for validation of scaling law analysis for the melt pool dimensions formed in LBPF single-track experiments in Ref. \cite{yang2022validated}. The transient heat transfer problem is solved in the domain containing powder layer and solid substrate (Fig. \ref{fem1}), both defined as continuum media with temperature-dependent and state-dependent properties. The main peculiarity of the model is an explicit definition of laser extinction in the volume of powder layer according to analytical solution of scattering/absorption problem in the powder media \cite{Gusarov2009}. This solution allows to define the volume heat source taking into account LPBF process parameters and powder properties. Also, this solution allows us to calculate the absorption efficiency of the powder-substrate system that was defined above as $A$. The governing equation of the model is the following:
\begin{equation}
\label{ge}
	\rho c \frac{\partial T}{\partial t} = \nabla\cdot (k \nabla T) + q + q^*
\end{equation}
where $T(\textbf x, t)$ is the temperature field that depends on spatial coordinates $\textbf x$ and time $t$; $q(\textbf x, t)$ is the volumetric moving heat source due to laser heat input; $q^*(\textbf x,t)$ is the additional volumetric heat source (sink) related to the latent heat of melting; $\rho$, $c$ and $k$ are the density, heat capacity and thermal conductivity of the media, respectively, that are defined depending on the phase state variable $\phi$ as follows\cite{Gusarov2009,yang2022validated,chernyshikhin2021selective}:
\begin{equation}
\label{prop}
\begin{aligned}
	k(\phi, T) &= \phi k_{a} +(1-\phi) k^{eff}_{pb}\\
	\rho c(\phi, T) &= \phi (\rho c)_a +(1-\phi) (\rho c)^{eff}_{pb}
\end{aligned}
\end{equation}
where $k_a$ and $(\rho c)_a$ are the properties of fused AlSi10Mg alloy, and $k^{eff}_{pb}$ and $(\rho c)^{eff}_{pb}$ are the effective properties of the powder bed. 

The initial values of state variable is set to zero $\phi=0$ (powder) in the powder bed layer, while in the substrate its initial value is $\phi=1$ (alloy).
 Irreversible change $\phi\rightarrow 1$ is then assumed for the powder bed, once its temperature becomes higher than the liquidus temperature, $T\geq T_l$ (see Fig. \ref{fem1}). 
 
 The temperature-dependent properties of the alloy are approximated based on the data of its known properties in the solid and liquid state (Table \ref{tab1}) as follows \cite{yang2022validated}:
\begin{equation}
\label{prop}
	k_{a} = 
		\begin{cases} 
  			k_s, \quad T<T_l\\
  			k_l, \quad T\geq T_l
		\end{cases},
	\qquad
	(\rho c)_{a} = 
		\begin{cases} 
  			\rho_s c_s, \quad T<T_l\\
  			\rho_l c_l, \quad T\geq T_l
		\end{cases}
\end{equation}

The effective properties of the powder bed are calculated based on the rule of mixture for $(\rho c)^{eff}_{pb}$ and by using semi-impirical ZSS model for $k_{pb}^{eff}$ \cite{yang2022validated}:
\begin{equation}
\label{propp}
\begin{aligned}
	(\rho c)^{eff}_{pb} &= f \rho_s c_s + (1-f) \rho_g c_g,\\
	k_{pb}^{eff} &= (1-\sqrt f) k_g + \sqrt f \frac{2}{1-B \bar k}K\\
	K = \tfrac{(1-\bar k)B}{(1-\bar k B)^2} &\log \tfrac{\bar k}{B} -\tfrac{B+1}{2}-\tfrac{B-1}{1-\bar k B},\qquad
	B = 1.25 \left(\tfrac{f}{1-f}\right)^{10/9}
\end{aligned}
\end{equation}
where $f$ is the experimentally measured packing density of the powder bed (Table \ref{gran}); $\rho_g$ $c_g$ and $k_g$ are the temperature dependent properties  of argon gas that were prescribed according to the tabulated data from material library of Comsol Multiphysics; $\bar k = k_g/k_s$ is the ratio between thermal conductivities of gas and solid material.

Volumetric heat source (sink) due to solid/liquid phase transition is defined in the heat balance equation \eqref{ge} as follows:
\begin{equation}
\label{qx}
\begin{aligned}
	q^* = 
	\begin{cases}
		-\rho L_m \frac{\partial \theta}{\partial t},  \quad T\geq T_l\\
		0, \quad T<T_l
	\end{cases}
\end{aligned}
\end{equation}
where $L_m$ is the latent heat of melting (Table \ref{tab1}), $\theta = \min\left (t/\Delta t,1 \right)$ is an auxiliary variable that grows linearly from 0 to 1 during the characteristic phase transition time $\Delta t$, then $\theta$ takes constant unit value, which consequently implies $\partial \theta/\partial t=0\Rightarrow q^*=0$. 

Generally, the characteristic time $\Delta t$ in \eqref{qx} can be chosen enough small to avoid its influence on the results of numerical simulations, although it should be non-zero to provide regularization of numerical solution. For simulations, it was estimated based on the relation $\Delta t = R^2/\alpha \approx 2.6\cdot 10^{-5}$ s with $\alpha=k_l/(\rho_l c_l)$, i.e. we used characteristic time for the heat diffusion in liquid alloy over the radius of laser beam. 

For the volumetric heat source $q(\textbf x, t)$ in the governing equation \eqref{ge} we prescribed the movement along the positive direction of $x$ axis with speed $v$ and nominal laser power $P$ according to the LPBF process parameters considered in the experiments.  This heat source is defined according to the solution of scattering/absorption model for interactions of Gaussian laser beam with the powder bed based on the following relation \cite{Gusarov2009}: 
\begin{equation}
\label{q}
\begin{aligned}
	&q = \begin{cases}
		\chi\, Q_0(x-vt,y) \tfrac{d Q(\xi)}{d\xi},  \quad z\geq-h\\
		0, \quad z<-h
	\end{cases}\\
	&Q_0(x,y) = \tfrac{P}{\pi R^2}\,\text {Exp}\left[{-\tfrac{r^2}{R^2}}\right],
	\quad \beta = \tfrac{3}{2}\tfrac{f}{1-f} \tfrac{\eta}{d_{50}}
\end{aligned}
\end{equation}
where $r=\sqrt{x^2+y^2}$ is the radial distance in $xy$ plane, $\chi$ [1/m] is an extinction coefficient, $\eta$ is the non-dimensional correction coefficient that takes into account the influence of PSD on the extinction coefficient (in the case $\eta=1$ the model corresponds to the powder bed consisted of spherical particles of the same size equals to $d_{50}$ \cite{Gusarov2009}); $Q_0(x,y)$ is the Gaussian distribution of laser power over the surface of powder bed; $Q(\xi)$ is the dimensionless function, which derivative defines the attenuation of laser power inside the powder bed and which depends on the dimensionless coordinate $\xi = \chi z$. 
Closed form representation for $Q(\xi)$ was obtained in Ref. \cite{Gusarov2009} based on the solution of integro-differential equation. This function depends on single physical property of the powder material -- its hemispherical reflectivity, that is set to be $R_s = 0.9$ that corresponds to the reflectivity of rough surface of aluminum alloy at the wavelength of Yb fiber laser \cite{bogdanova2024mesoscale}. To make this paper self-contained, solution for $Q(\xi)$ is presented in Appendix A.

The numerical simulations were performed for the one-half of the domain taking into account symmetry of the problem with respect to the plane $y=0$ (see Fig. \ref{fem1}). 
On the lateral boundaries of the domain we used the heat insulation conditions ($\nabla T\cdot\textbf n=0$). At the lower boundary we prescribed constant temperature condition $T=T_0$, while at the upper boundary we took into account the convective and radiative heat losses:
\begin{equation}
\label{bc}
\begin{aligned}
	z=0: \qquad -k \nabla T\cdot\textbf n = h_c(T-T_0)+\varepsilon\sigma (T^4-T_0^4)
\end{aligned}
\end{equation}
where $\textbf n$ is outer unit normal at the boundary, $h_c$ is the temperature-dependent heat transfer coefficient that was estimated based on the known semi-empirical assessments for the natural convection parameters over the hot horizontal surface (argon was not blown in the considered short-time single track experiments), the value of this coefficient reached maximum value $h_c\approx$ 6.5 W/($m^2$K); $\varepsilon=A_s=1 - R_s= 0.1$ is emissivity of the surface that is assessed based on the prescribed value of reflectively $R_s$ and Kirchhoff's law; $\sigma$ is Stefan-Boltzmann constant.

The size of the domain in numerical simulations was chosen enough large to avoid the influence of lateral and lower boundaries on the thermal state in the fused zone and to avoid the influence of transient effects on the predicted dimensions of the melt pool (i.e. of the single track). These dimensions of melt pool were defined as the size of the area, where the sate variable takes unit value $\phi=1$ (see Fig. \ref{fem1}). 

\begin{figure}[b!]
 \centering
  (a)\includegraphics[width=0.45\linewidth]{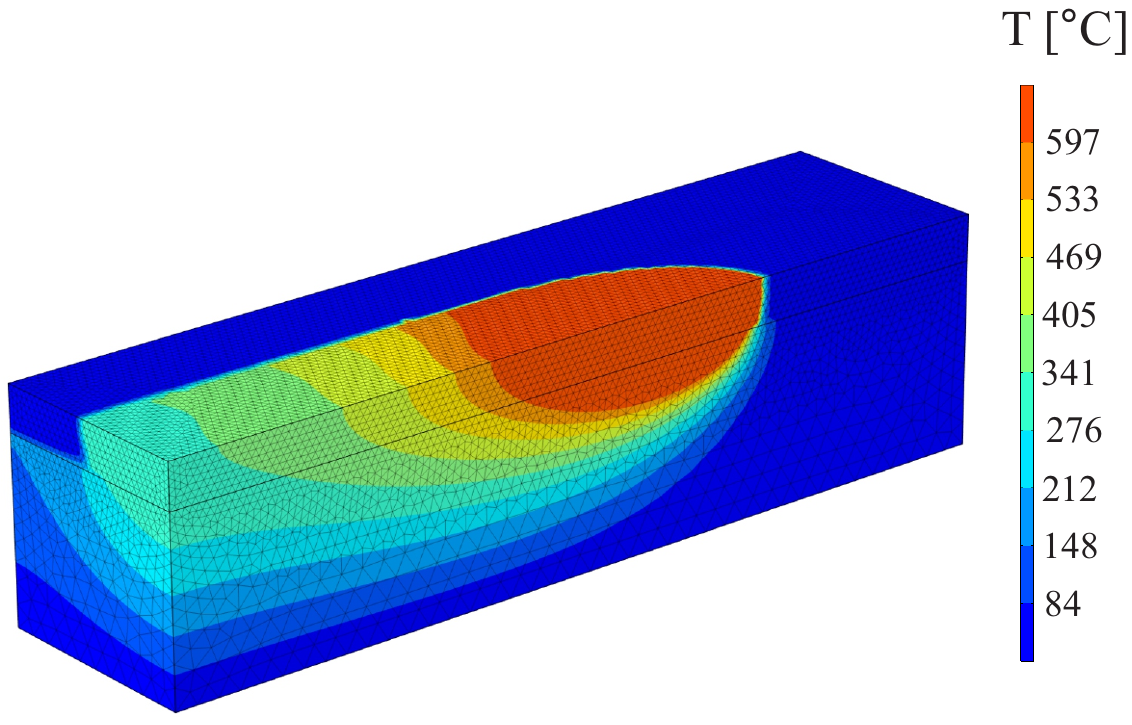}\qquad
  (b)\,\,\includegraphics[width=0.2\linewidth]{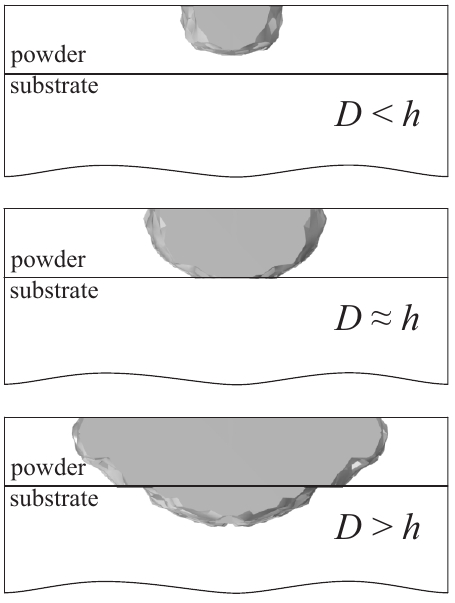}
  \caption{Examples of the results of numerical simulations, (a): thermal state of the powder layer/substrate system ($P=300$ W, $v=1.745$ m/s, time moment $t=4.5$ ms), (b): the shape of melt pool cross sections that arise for different intensity of heat source. The position of lack of fusion boundary in numerical simulations is attributed to the case of process parameters ($v,P$) that provides $D\approx h$}
  \label{fem2}
\end{figure}

 The model was implemented in Comsol Multiphysics. From numerical simulations we found the distribution of temperature field $T(\textbf x, t)$ and state variable $\phi(\textbf x, t)$ in the domain under prescribed action of moving heat source. The example of such solutions for temperature field is given in Fig. \ref{fem2}a and for the melt pool dimensions in Fig. \ref{fem2}b. The objective of the numerical simulations was to predict the lack-of-fusion boundary on the experimental process map by identifying the process parameters, at which the melt pool depth $D$ equals the powder layer thickness $h$ (see Fig. \ref{fem2}b). Notably, that considered model \eqref{ge}-\eqref{bc} does not contain absorptivity $A$ as the parameter. Instead, the model explicitly defines the attenuation of laser power in the powder bed according to Eq. \eqref{q}. In this relation, we have single unknown parameter that should be evaluated based on the experimental data -- the correction coefficient $\eta$ that defines the influence of particle size distribution on the extinction coefficient $\chi$. This correction coefficient was identified recently for LPBF of 316l stainless steel in Ref. \cite{yang2022validated} based on the analysis of experimentally measured  melt pool width. It was found that with increase of the laser beam diameter from 66 $\mu$m to 196 $\mu$m, the correction coefficient ranges in $0.85...0.47$. In the present study, we identify this coefficient $\eta$ based on the analysis of the position of lack of fusion boundary. Namely, in numerical simulations, we used the following algorithm for each type of the   powders P0...P5:

1. We began by considering the maximum laser speed employed in the experiments 
($v^* = 1.98$ m/s) and the corresponding critical laser power $P^*$ that defines the lack-of-fusion boundary on the experimental process map of single-track experiments. For these parameters, we performed a series of simulations while varying the correction coefficient $\eta$ to identify the specific value at which the predicted melt pool depth matched the powder layer thickness $D\approx h$ (Fig. \ref{fem2}b) -- that is, the condition where the single track is fully fused to the substrate. In such a way, we identified the value of $\eta$ for a given type of the powder.

2. Using the found value of $\eta$, we performed simulations for the all other  combinations of critical speed and power $(v^*,P^*)$ corresponding to the lack of fusion boundary. This allowed us to evaluate the range of validity of the model against experimental data for a given type of the powder and different values of scanning speed (and related Peclet numbers).

3. We calculated the absorptivity of the powder/substrate system $A$, that can be done analytically by using the found value of $\eta$ (and the corresponding value of extinction coefficient $\chi$) and the solution for attenuation function $Q(\xi)$ \eqref{cQ}. Namely, the ratio between the absorbed and the incident laser power within the considered volumetric heat source model can be estimated as:
\begin{equation}
\label{afem}
\begin{aligned}
	A = Q(0) - Q(-\lambda) 
\end{aligned}
\end{equation}
where $\lambda = \chi h$ is the optical thickness of powder layer.

\subsubsection{Eager-Tsai model for the melt pool width}

In Ref. \cite{doring2025trans} authors developed and validated an approach for absorptivity evaluation within the LPBF processes and semi-analytical calculations performed based on the Eager-Tsai model (ETM) \cite{eagar1983temperature}. Since this model can accurately predict the melt pool width within the wide range of process parameters (in contrast to the melt pool depth that strongly deviates from ETM predictions for high energy regimes \cite{doring2025trans}), it was suggested to fit the solution of ETM to the experimentally measured melt pool width by using absorptivity coefficient $A$ assuming that all other material and process parameters are known. Thus, we use this method in the present study and perform the fitting of ETM solution that defines the following quasi-stationary (in the moving frame) distribution of the temperature field over the surface of half space due to the action of moving Gaussian heat source \cite{eagar1983temperature,gladush2011physics,doring2025trans}:
\begin{equation}
\label{etm}
\begin{aligned}
	T(\bar x,\bar y,\bar z) = \frac{T_l\,\beta}{\sqrt\pi} \int\limits_0^\infty
	\frac{
	\,\text {Exp}\left[{
			-\frac{\bar z^2}{\bar t^2}
			-\frac{(\bar x+\bar t)^2}{8(1+4\bar t/Pe)}
			}\right]
		}
		{
		\sqrt {\bar t\,} (1+4 \bar t/Pe)
		} \,d\bar t
\end{aligned}
\end{equation}
where it is assumed that origin of coordinate system is located at the center of moving heat source on the surface of half space; this heat source moves along the positive direction of $x$-axis; $z$-axis is oriented in the outer direction of half space; $\beta$ is given by Eq. \eqref{b}, and the dimensionless coordinates and time are introduced as follows: $\bar x= 4x/\omega$, $\bar y= 4y/\omega$, $\bar y= 4y/\omega$, $\bar z= 16\sqrt{2}z/(\omega\sqrt {Pe})$ and $\bar t = t \omega/(4v)$. 

For given values of Peclet number $Pe$ and normalized ethalpy $\beta$ we can calculate the distribution of temperature field by using \eqref{etm} and find an isotherm that corresponds to the solidus temperature $T=T_s$. Maximum width of this isotherm in the transverse direction (along $y$-axis at $z=0$) defines the width of the melt pool $W$. These calculations were performed in symbolic algebra system by using numerical integration. Properties of liquid AlSi10Mg alloy were used (Table 2). The values of absorptivity coefficient $A$ were found based on fitting of ETM predictions for melt pool width $W$ to experimental data for the width of single tracks obtained based on microscopy analysis of cross section images.

\section{Results and Discussion}
\subsection{Single track experiments}

\begin{figure}[b!]
\centering
  \includegraphics[width=0.9\linewidth]{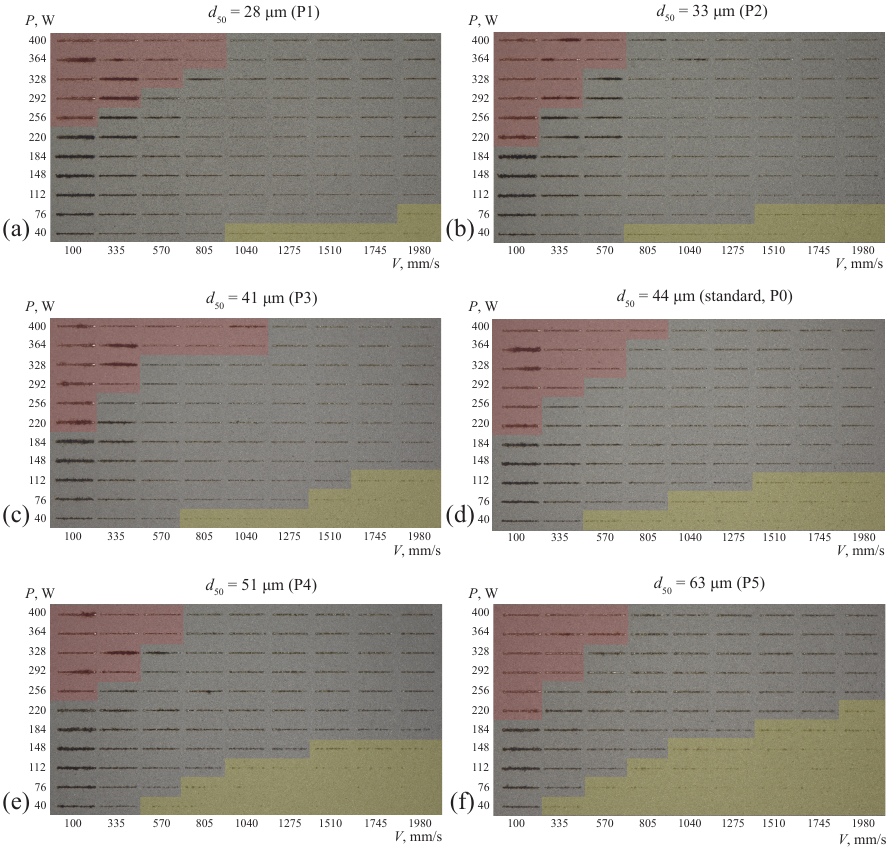}
  \caption{Top view of all single tracks obtained for different powders, (a): P1, (b): P2, (c): P3, (d): P0, (e): P4, (f): P5. The images are ordered with respect to the mean size of powder particles that is noted over the images. Red color corresponds to the keyhole regime of melting. Yellow color defines the lack of fusion zone, i.e. the single tracks with irregular and strongly discontinuous structure or absence of tracks}
  \label{fig5}
\end{figure}

The top views of all single tracks produced by using different powders P0-P1 are presented in Fig. \ref{fig5}. In these plots, we highlight with red color the tracks that correspond to the keyhole mode of melting, while the lack of fusion regions (discontinuous tracks with strong balling effects, or the absence of tracks) are shown by yellow color. 
The assessment for the keyhole mode was carried out based on the analysis of cross-section micrographs. The tracks with a depth-to-width ratio exceeding $D/W>$0.75 are highlighted in red in Fig. \ref{fig5}.
The evaluation of the lack of fusion zone was performed by analyzing the magnified top-view images. Discontinuous tracks that were fused less than 50\% along their length are highlighted in yellow in Fig. \ref{fig5}. Thus, we distinguish the approximate transition (lack of fusion) boundary between tracks whose depth was equal to or greater than the layer thickness ($D \geq h$) and tracks with insufficient depth for bonding with the substrate ($D < h$).

 In Fig. \ref{fig5}, it can be seen that the keyhole zone is almost independent on the powder type and it is localized in the region of laser power above $P=$ 220 W and speeds below $v=$ 0.8 m/s for all powders P0-P5. At the same time, the size of the lack of fusion zone significantly depends on the powder type. For  the powders with smaller particle size, this zone becomes smaller.
For powder P1 ($d_{50}=28\,\mu$m, Fig. \ref{fig5}a), almost all tracks appear fully fused, and only upon closer examination of magnified images we could identify tracks that meet our criterion for the lack of fusion boundary (less than 50\% fusion along the length). 
For powders with a larger mean particle size, the lack of fusion zone increases, and for the coarsest powders considered, it occupies almost half of the process map (Fig. \ref{fig5}f).
For the standard powder P0, which $d_{50}$ value is approximately in the middle of all considered fractions, the lack of fusion zone is also of intermediate size -- it is localized at the laser power levels below $P=$ 120 W and speeds above $v=$ 0.57 m/s. Thus, Fig. \ref{fig5} demonstrates a significant effect of powder particle size on the lack of fusion zone, which shifts toward lower speeds and higher laser power zones for coarser fractions of AlSi10Mg powder.

\begin{figure}[b!]
\centering
  \includegraphics[width=0.9\linewidth]{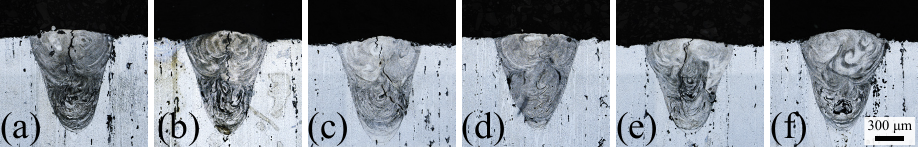}
  \caption{Microscopy of single tracks cross sections that were obtained by using LPBF process regimes with $P=328$ W and $v=0.335$ m/s (keyhole mode) for  different powders (a): P0, (b): P1, (c): P2, (d): P3, (e): P4, (f): P5}
  \label{fig6}
\end{figure}
\begin{figure}[b!]
\centering
  \includegraphics[width=0.7\linewidth]{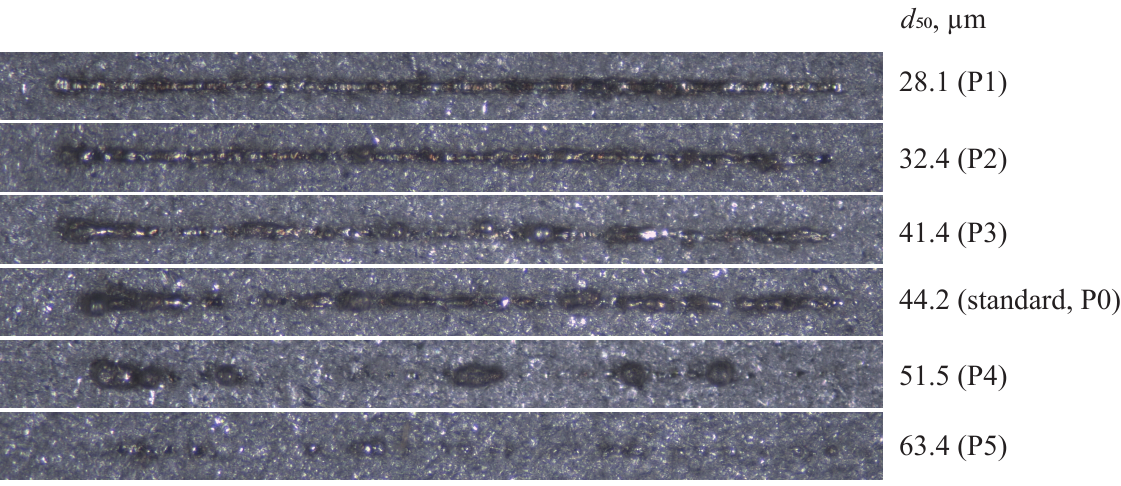}
  \caption{Magnified top view of single tracks obtained by using LPBF process regimes with $P=148$ W and $v=1.51$ m/s for different powders P0-P5}
  \label{fig7}
\end{figure}

The examples of cross section images and magnified top view photographs of single tracks are presented in Figs. \ref{fig6}, \ref{fig7}. In Fig. \ref{fig6} we compare the melt pools formed in the keyhole regime ($P=328$ W and $v=0.335$ m/s) using different powder types P0-P5. It can be seen that the cross-section geometry of these melt pools remains nearly identical, indicating that the mean powder size has negligible influence on melting behavior under high input energy conditions.
Figure 7 presents the magnified top-view images of tracks produced at power $P=148$ W and speed $v=1.51$ m/s using different powders P0-P5. These process parameters lie close to the lack of fusion boundaries. However, while fine powders P0-P3 can be attributed to the fusion zone ($D\geq h$ almost over the whole length of track), the use of coarse powders P4, P5 result in strongly discontinuous tracks that correspond to the lack of fusion zone (marked yellow in Fig. \ref{fig5}). 

\begin{figure}[b!]
\centering
  (a)\includegraphics[width=0.45\linewidth]{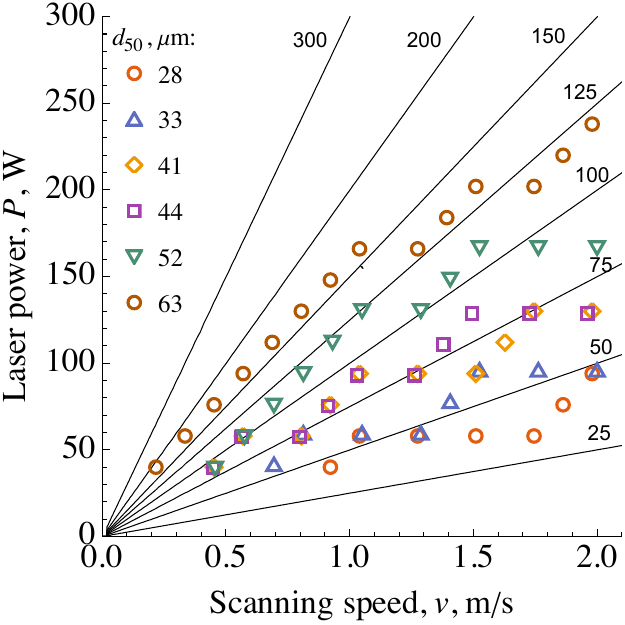}\,\,\
  (b)\includegraphics[width=0.45\linewidth]{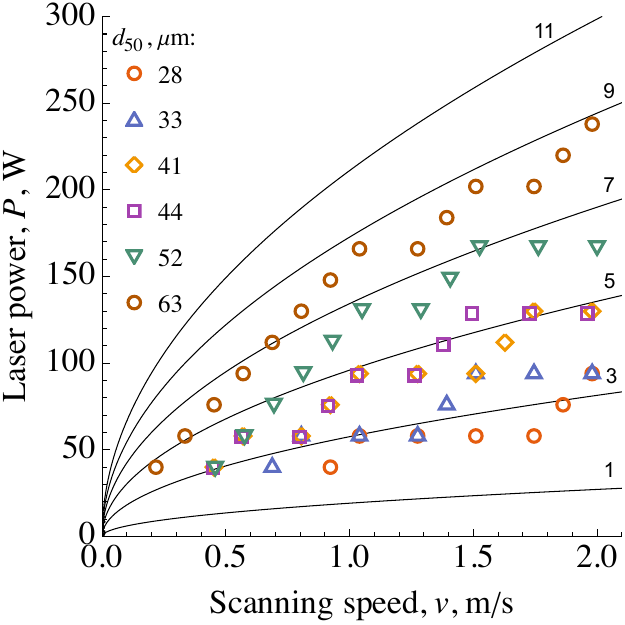}
  \caption{The lack of fusion boundaries on the process map of single track experiments performed with different type of powders P0-P5. Different dots corresponds to the experiments with powders having different size of particles $d_{50}$ (noted on the plots). The comparison is presented with the contour lines of constant linear energy density $E_l$ [J/m] (a) and with contour lines of constant enthalpy density $\Delta H$ [J/mm$^3$] (b)}
  \label{fig8}
\end{figure}

\subsection{Scaling law analysis}
Next, we can evaluate the correspondence between the experimentally determined lack of fusion boundaries and theoretical contours that define constant linear energy density $E_l=P/v$ and constant enthalpy density $\beta$ \eqref{b} on the process map. This analysis is shown in Figure 8. For clarity, in these figures (and in subsequent similar figures), we represent the experimental lack of fusion boundaries as points whose coordinates were determined by averaging the power and scan speed between adjacent tracks that fell into different zones (fused/lack of fusion) presented in Figure 5 for each type of the powder P0-P5. 
Plots in Fig. 8a and 8b show the comparison of these experimental boundaries with contours of constant linear energy density and constant enthalpy density, respectively. 
It can be seen that at low laser scan speeds, the boundaries show good agreement with constant linear energy density contours (Fig. 8a). However, for speeds $v > 1...1.2$ m/s, they deviate from these contours toward lower values of laser power.
At the same time, in the high-speed regimes, the experimental data points demonstrate better agreement with contour lines corresponding to constant enthalpy density (Fig. 8b).
For fine powders (P1, P2), this transition between criteria is less pronounced. However, for other powders, including the standard AlSi10Mg powder, it is clearly visible in the presented graphs.
Therefore, in our theoretical analysis, we separately considered these distinct parameter zones with the use of corresponding different criteria for the lack of fusion boundary. 

\begin{figure}[b!]
\centering
  (a)\includegraphics[width=0.45\linewidth]{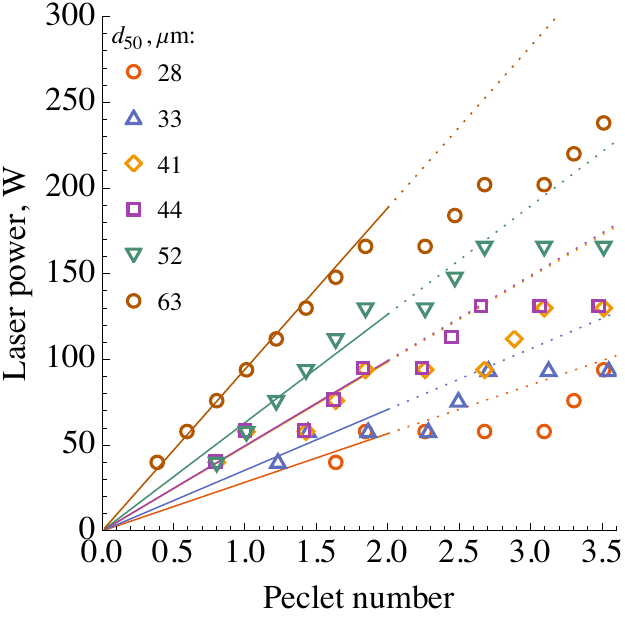}\,\,\
  (b)\includegraphics[width=0.45\linewidth]{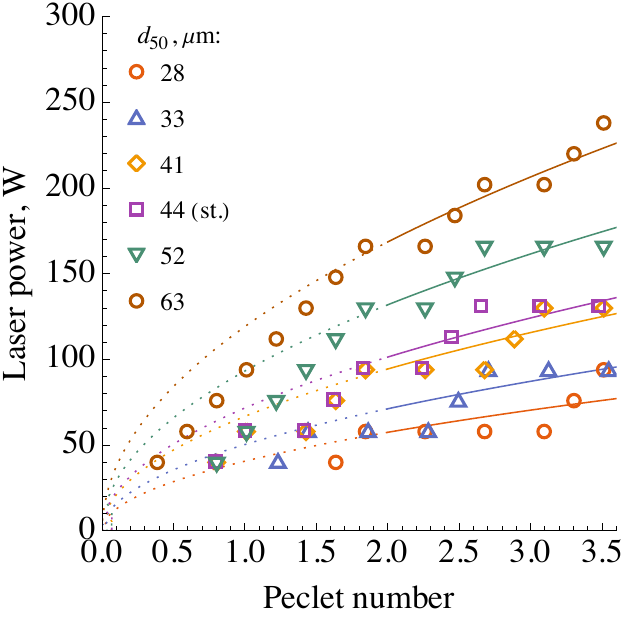}
  \caption{Best fit of theoretical criteria for the lack of fusion boundary (lines) to the experimental boundaries (dots) obtained for the powders P0-P5 with different mean size of particles, (a): criterion for the single track melting efficiency ($N=1$, Eq.\eqref{n}), (b): criterion for the normalized enthalpy density $(\beta=1$, Eq.\eqref{b})}
  \label{fig9}
\end{figure}

The best fit of energy-based criterion ($N=1$, Eq. \eqref{n}) and temperature-based criterion ($\beta=1$, Eq. \eqref{b}) to the experimentally determined lack of fusion boundaries for different types of powders is presented in Fig. \ref{fig9}. In these plots, we use the dimensionless Peclet number ($Pe$) instead of scanning speed ($v$) on the abscissa axis to analyze the applicability of the criteria from a dimensionless analysis perspective. The correspondence between theoretical and experimental boundaries was established by adjusting the absorptivity coefficient $A$, as described in Section 2.3.1.
For the energy-based  criterion \eqref{n}, the analysis was performed for Peclet numbers $Pe<$ 2 (Fig. 9a), while for the temperature-based criterion \eqref{b} it was done for $Pe>$ 2 (Fig. 9b). Figure 9 shows that this particular Peclet number value $Pe = 2$ ($v =$ 1.13 m/s) proves critical in terms of matching the experimental boundary with either constant energy density or constant enthalpy density.
The corresponding values of absorptivity $A$ determined using both criteria through equations \eqref{a} and \eqref{ab}, along with critical linear energy density and critical enthalpy density for all powders are presented in Table 3 and Figure 10.

\begin{table}[t!]
\centering
\footnotesize
\caption{Identified values of AlSi10Mg powders absorptivity based on the scaling law analysis (SLA) and finite-element simulations (FEM)}
\begin{tabular}{ccccccccccc}
\hline
Powder & $d_{50}$ & $E^*_l$& $\Delta H^*$& \multicolumn{3}{c}{Absorptivity, $A$} \\
type &  ($\mu$m) & (J/m) &(J/mm$^3$)& SLA ($\text{Pe}<2$) & SLA ($\text{Pe}>2$) & FEM \\
\hline
P0 & 44.2 &88&8.9& 0.22 & 0.20 & 0.26\\
\hline
P1 & 28.1 &50&4.4& 0.38 & 0.35 & 0.37 \\
\hline
P2 & 32.8 &63&6.0& 0.31 & 0.28 & 0.33 \\
\hline
P3 & 41.4 &88&7.8& 0.22 & 0.22 & 0.27 \\
\hline
P4 & 51.5 &112&11.7& 0.17 & 0.15 & 0.22 \\
\hline
P5 & 63.4 &167&14.6& 0.11 & 0.12 & 0.16 \\
\hline
\end{tabular}
\label{Ap}
\end{table}

\begin{figure}[t!]
\centering
  (a)\includegraphics[width=0.45\linewidth]{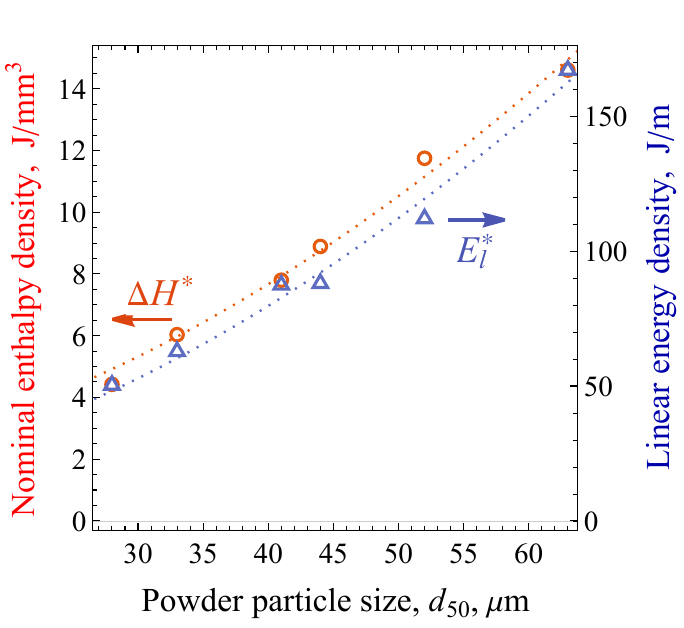}
  (b)\includegraphics[width=0.42\linewidth]{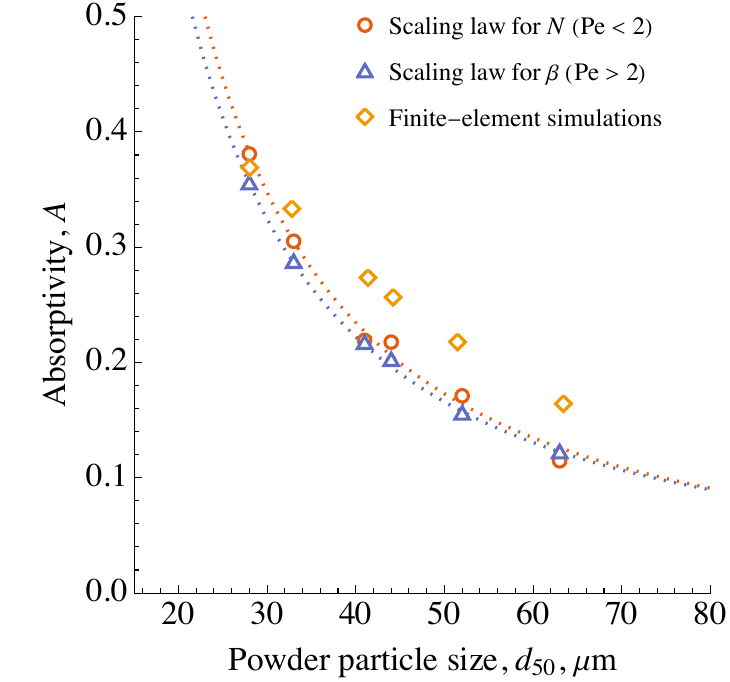}
  \caption{Established dependences of critical linear energy density and critical enthalpy density (a) and absorptivity (b) on the powder particle size. Trend lines are shown by dotted lines}
  \label{fig10}
\end{figure}

The performed scaling law analysis revealed that for the investigated powders, the critical energy density values range between 50-167 J/m, while the critical enthalpy ranges between 4.4-14.6 J/mm$^3$ (see Table 3). These values sohuld be treated as nominal, since they were calculated based on the process parameters ($P,v$) without accounting for the absorption coefficient.
Notably, smaller powder particles require less energy to form stable tracks (Fig. 10a). We attribute this effect to variations in the absorptivity among powders with different particle sizes -- coarser powders exhibit reduced absorption (Fig. 10b). As it was discussed in the Introduction, this effect is typical for LPBF processes, and in the present case we have quantitatively confirmed it through the scaling law analysis for the single track experiments performed with different fractions of AlSi10Mg powder.
The determined range of absorptivity values is typical for aluminum powders, though as noted in the Introduction, literature data show considerable variability for this property. It is interesting to note that applying two distinct solutoins (2) and (4) for different Peclet number ranges yields remarkably consistent values of absorptivity $A$ (see Fig. 10b, red and blue points). It is also worth noting that the absorptivity of the coarsest powders approaches the absorptivity of the particles material that is approximately $A_s=1-R_s=0.1$ (it was also used in the numerical simulations, see Section 2.3.2) \cite{bogdanova2024mesoscale}.

The trend lines in Fig. 10 were defined using a dimensionless parameter that is usually introduced for the description of powder particle size effects in LPBF processes: $\bar d = d_{50}/(2\omega$) \cite{oliveira2020revisiting}. In particular, the following functions were used to approximate the established dependencies for critical enthalpy density and linear energy density:
\begin{equation}
\label{regr}
\begin{aligned}
	\Delta H^* =H_s + 37.2 \,\bar d^{\,1.78},\qquad
	E_{l}^* = E_s + 423\,\bar d^{\,1.76}
\end{aligned}
\end{equation}
where $H_s=1.798$ J/mm$^3$ is the melting enthalpy of solid AlSi10Mg alloy, and the corresponding linear energy density is calculated as $E_s = \omega hH_s = $ 12.2 J/m.

For the absorptivity we used the following function to plot the trend lines in Fig. 10b:
\begin{equation}
\label{regr}
\begin{aligned}
	A = a\,\bar d^{\,-b}
\end{aligned}
\end{equation}
in which $a=0.0566$, $b = 1.367$ for the energy-based scaling low for $N$ (red line in Fig 10b) and $a=0.0567$, $b = 1.312$ for the temperature-based criterion for $\beta$ (blue line in Fig 10b).

\begin{figure}[b!]
\centering
  (a)\includegraphics[width=0.45\linewidth]{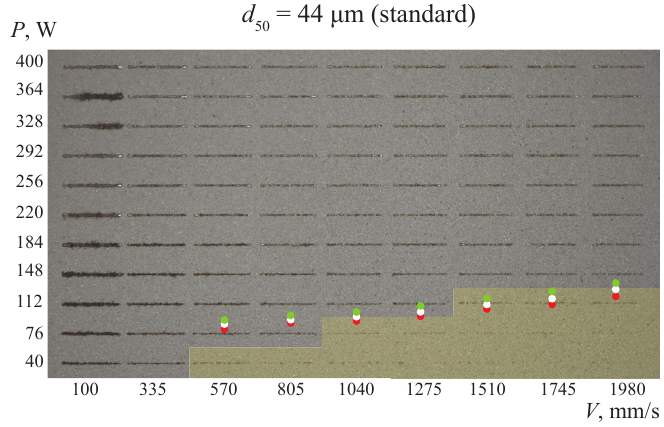}
  (b)\includegraphics[width=0.45\linewidth]{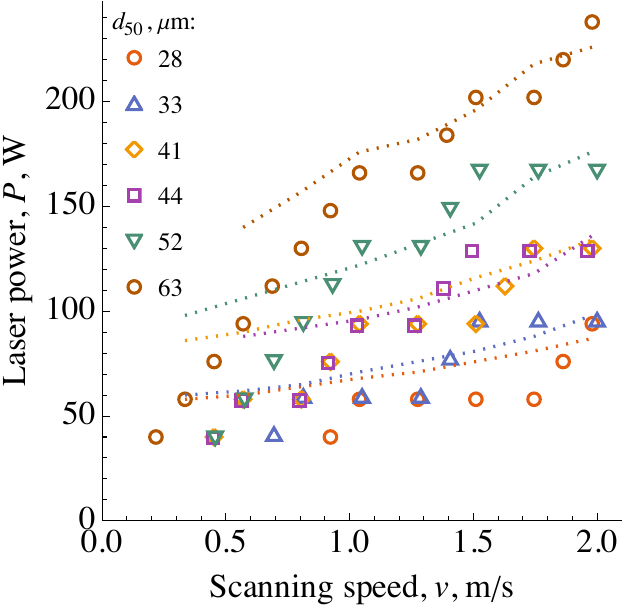}
  \caption{Comparison of experimental data and the results of numerical simulations, 
(a): Effect of the correction factor on the position of predicted lack of fusion boundary for standard powder P0 (green points: $\eta$ = 0.7, white points: $\eta$ = 0.75, red points: $\eta$ = 0.8),
(b) Comparison between experimental and theoretical lack of fusion boundaries for all powder types (P0-P5), with corresponding mean particle sizes indicated in the plot}
  \label{fig11}
\end{figure}

\subsection{Results of numerical simulations}

The results of numerical simulations are presented in Fig. 11. Figure 11a illustrates the selection of the correction coefficient $\eta$ in the moving heat source model \eqref{q}, which determines the influence of PSD on the extinction coefficient $\beta$ (see Section 2.3.2). This illustration is given for the standard powder P0 as an example. Figure 11a shows that the higher values of the correction coefficient (and, accordingly, the higher extinction coefficient) shifts the lack of fusion boundary downward -- toward the zone of lower energy input. 
Thus, by varying this coefficient, we determined its value for all types of powders P0-P5. A comparison between the all obtained theoretical and experimental lack of fusion boundaries is presented in Fig. 11b. In Appendix A we also list  the identified values of the correction factor $\eta$, extinction coefficient $\chi$, as well as superimposed numerical simulation results on the all top-view photographs of single tracks. 

In Fig. 11b it is seen that the numerical modelling agrees well with the experiment in parameter regions where the laser speeds are sufficiently high. Similar to the scaling law analysis, we can conclude here that the numerical model is accurate in the range of Peclet numbers Pe $> 2$ ($v > 1.1$ m/s).
It could be assumed that if we had used a different approach and calibrated the correction factor in the low-speed region, the agreement would have been achieved there as well. However, this is not the case because the numerical model fundamentally fails to capture the geometry of the experimentally obtained lack of fusion boundaries at low speeds and significantly overestimates the critical level of input energy (see Fig. 11b, Fig. A.1). This is due to the fact that the applied moving heat source model is based on the assumption of laser-powder particle interaction, which only holds true at high Peclet numbers. At low Peclet numbers, the interaction occurs primarily with the melt pool surface without strong scattering effects in the powder media, making the considered numerical model invalid.

It is also worth noting that good agreement between the experimental and numerical results in Fig. 11b was achieved for all powders except the finest one (P1). For this powder, the highest extinction coefficient was obtained (see Appendix A), yet even this elevated value fails to describe the experimental data.
In fact, to accurately describe the experiment for P1 within the existing model, we would need to introduce additional assumptions -- for instance, a dependence of the extinction coefficient on the Peclet number or consider the generalized constitutive relations for powder media \cite{solyaev2022gradient}. This, however, would require further investigation and the development of a more complex numerical models.

The absorptivity values $A$, calculated using the obtained extinction coefficient (Fig. A.2c) and formula (12), are presented in Fig. 10b (yellow dots) and listed in Table 2. It can be seen that numerical simulations also validate the particle size effect on absorptivity, although the $A$ values derived based on equation \eqref{afem} are somewhat higher than those determined by scaling law analysis for all powders (except for the finest one, for which the numerical model is less accurate).
The refinement provided by numerical modelling is associated, in particular, with accounting for the temperature and phase-state dependence of the powder and substrate properties. Therefore, the estimates obtained from scaling law analysis can be considered as the lower bound for the absorptivity.

\subsection{Evaluation of absorptivity based on single track width analysis and ETM}

\begin{figure}[t!]
\centering
  (a)\includegraphics[width=0.45\linewidth]{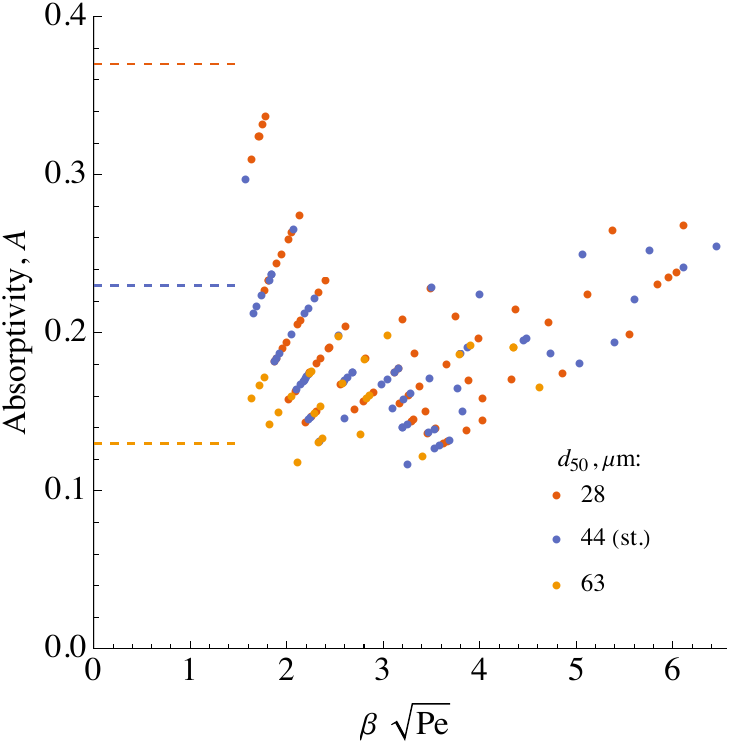}
  (b)\includegraphics[width=0.45\linewidth]{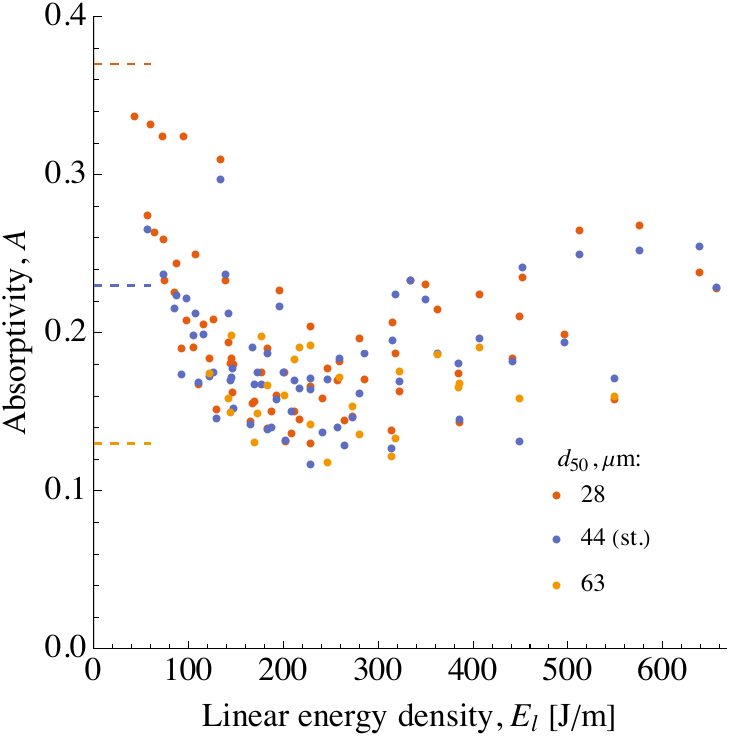}
  \caption{Identified dependence of absorptivity on $\beta\sqrt {Pe}$ (a) and on $E_l$ (b) for three types of powders (P0, P1, P5) based on the analysis of single track width and ETM. Horizontal dashed lines correspond to the absorptivity values evaluated for these powder based on the analysis of position of the lack of fusion boundary (according to Table 3)}
  \label{fig12}
\end{figure}

The results of processing of the experimental data on the melt pool width using the ETM model (Section 2.3.3) are presented in Figs. 12, 13.
Fig. 12 shows the identified absorptivity for three powder types: the finest (P1), standard (P0), and the coarsest (P5). These powders were selected as the most representative examples, while others were omitted to avoid cluttering the figure. Each data point in Fig. 12 corresponds to a single track produced under specific combination of printing parameters $(P, v)$, for which the experimental melt pool width was matched with the ETM model predictions to determine $A$.
In Fig. 12a, the data points are arranged according to the normalized enthalpy multiplied by the square root of the Peclet number ($\beta\sqrt {Pe}$), as proposed in the original study where this identification method was developed \cite{doring2025trans}. For clarity, in Fig. 12b we also present the derived dependence of $A$ on the linear energy density. A typical trend for $A$ is observed. At high energy inputs ($\beta\sqrt {Pe} > 3$ and $El > 200$ J/m), $A$ exhibits a monotonic increase, attributed to the transition to keyhole-mode of melting \cite{doring2025trans}.
At low energy inputs, $A$ also increases, particularly pronounced for the fine powder P1. This effect is observed for other powders as well, though to a lesser extent.
The rise in $A$ at low energy inputs aligns with estimates derived from our previous analysis. 
Namely, the horizontal dashed lines in Fig. 12 correspond to the values of $A$ determined from the analysis for the lack of fusion boundary positions in top-view images of single tracks. Specifically, to construct these lines, we used the average values of $A$ obtained from the three methods we applied (SLA and FEM, see Table 2). It can be observed that the variation of the absorption coefficient $A$ derived from the melt track width analysis generally aligns with our earlier estimates at low levels of input energy. In this region of parameters -- which is closest to the lack of fusion boundary -- the coarsest powder shows the smallest increase in $A$ (not exceeding 0.22), while the finest powder (P1) exhibits growth up to $A \approx 0.35$. The standard powder (P0) also demonstrates an increase in $A$ at low energy inputs, though less pronounced than that of the fine powder.

Thus, we observe a particle size effect on absorptivity $A$ that becomes pronounced at low energy inputs. This effect could potentially be utilized to optimize LPBF processes for highly reflective/conductive alloys, for which the improvement of laser absorption remains essential for producing fully dense materials, e.g. for the advanced thermal management systems \cite{sefene2022state,solyaev2019overmelting,chen2023high}.
For instance, one could employ fine powder fractions and conduct LPBF using low levels of input energy (in terms of $E_l$ or $\Delta H$) while maintaining elevated platform temperature ($T_0$). As follows from equations \eqref{a}, \eqref{ab}, this approach would be equivalent to using higher laser powers, while simultaneously enabling enhanced material absorptivity through the particle size effects that is observed in the present study.
Other process parameters would obviously require adjustment as well, including:
hatch spacing, scan strategies, powder layer thickness and potentially the laser spot diameter and wavelength, which will induce additional size effects on absorptivity \cite{doring2025trans,ye2019energy}.

\begin{figure}[b!]
\centering
  (a)\includegraphics[width=0.28\linewidth]{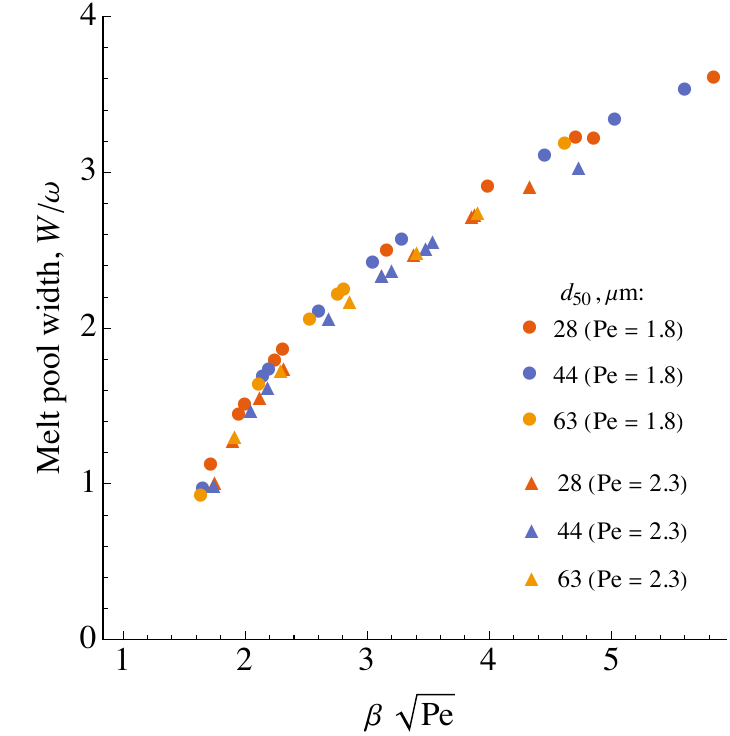}
  (b)\includegraphics[width=0.28\linewidth]{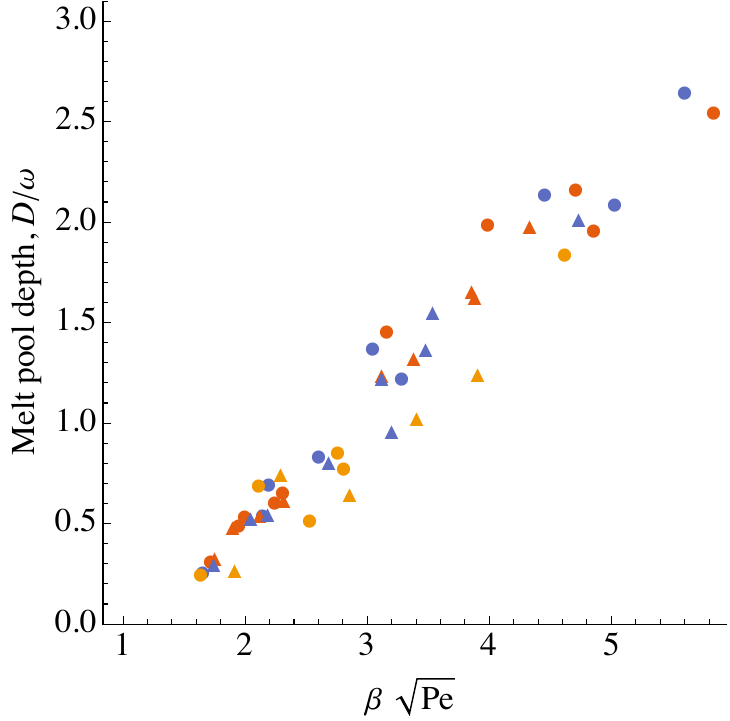}
  (c)\includegraphics[width=0.28\linewidth]{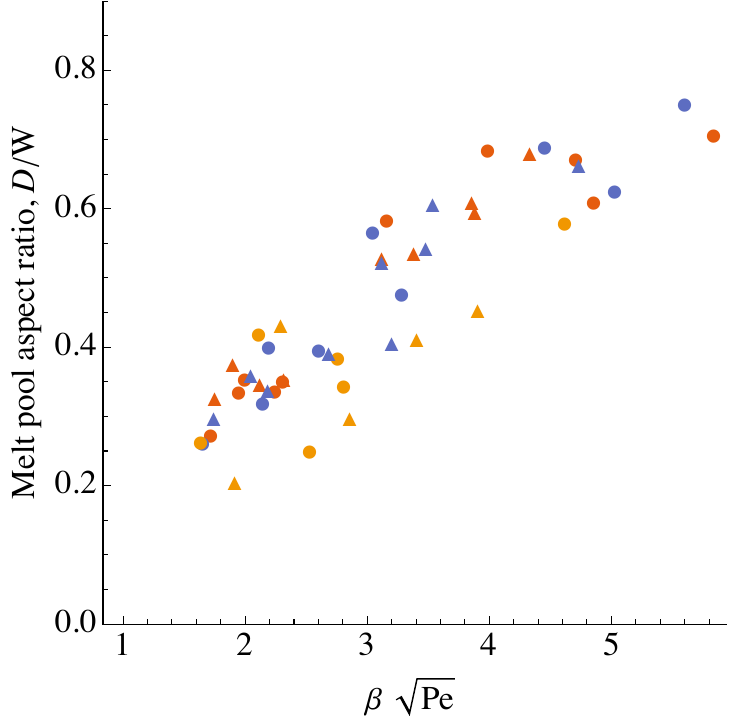}
  \caption{Established dependence of the melt pool normalized width (a), depth (b) and aspect ratio (c) on $\beta\sqrt {Pe}$ based on the analysis of single track width and ETM for the powders with different mean size of particles and at different Peclet numbers (noted on the plot)}
  \label{fig13}
\end{figure}

The illustration for the investigated melt pool dimensions with the use of ETM is presented in Fig. 13. In this analysis, the absorptivity adjustment ensures precise matching between ETM predictions and experimental dependencies of the melt pool width on $\beta\sqrt {Pe}$ (Fig. 13a) (this parameter $\beta\sqrt {Pe}$ was alternatively referred to as the "operation parameter" in the original work \cite{eagar1983temperature}).
Consequently, all experimental data points in Fig. 13a align with smooth, monotonically increasing curves corresponding to ETM solutions at given Peclet numbers. For illustration purposes, Fig. 13a presents results for two specific Pe values (1.8 and 2.3), resulting in two characteristic curves -- the ETM solutions for these Pe values.
However, such smooth dependencies do not emerge for melt pool depth and aspect ratio (Figs. 13 b, c), which stems from additional phenomena (hydrodynamic effects, vaporization, etc.) not accounted for in ETM. This limitation nevertheless doesn't restrict the model's applicability for the analysis of single track width \cite{doring2025trans}.
The results in Figs. 13 b,c reveal some moderate particle size effect: melt pool depth and aspect ratio are generally smaller when using coarse powder compared to standard and fine fractions. This powder size effect on melt pool geometry, while less pronounced than the previously observed effect on absorptivity ($A$), remains measurable, particularly for the coarsest powder fractions.

\subsection{Processing of known experimental data for various AlSi10Mg powders}

\begin{table}[b!]
\hspace{-1cm}
\footnotesize
\caption{Identified values of absorptivity (at low energy input) for standard AlSi10Mg powders based on the scaling law analysis and published experimental data}
\hspace{-2cm}
\begin{tabular}{cccccccccccccc}
\hline
\#&$d_{10}$ & $d_{90}$ & $h$& $2R$& $T_0$&$P^*$& $v^*$& $E^*_l$& $\Delta H^*$& $Pe$&\multicolumn{2}{c}{Absorptivity, $A$}& \\
&($\mu$m) &  ($\mu$m) &  ($\mu$m)&  ($\mu$m) & ($^o$C)& (W) & (m/s)& (J/m) &(J/mm$^3$) & (1) & Eq. (2) & Eq. (4) & Ref. \\
\hline
1&15 & 45 &  30& 150& RT$^*$ & 170&0.9& 189& 6.5 & 3 &0.09& \textbf{0.28}& \cite{kempen2015processing} \\
\hline
2&25 & 70 &  50& 100& 100& 60&0.8& 75& 4.5 & 1.8 &\textbf{0.24}& 0.37& \cite{aversa2018single} \\
\hline
3&25 & 70 &  50& 100& 100& 130&1.9& 68& 6.3 & 4.2 &0.32& \textbf{0.26}& \cite{aversa2018single} \\
\hline
4&15 & 53 &  30& 70& RT &130&1.1& 118& 5 & 3.4 &0.14& \textbf{0.36}& \cite{dong2020research} \\
\hline
5&20 & 63 &  400& 20& RT &100&0.75& 133& 86 & 0.3 &\textbf{0.24}& 0.02& \cite{aboulkhair2016formation} \\
\hline
6&33 & 70 &  60& 80& RT &58&0.57& 102& 7.2 & 1 &\textbf{0.19}& 0.25& this study 
\\\hline
7&33 & 70 &  60& 80& RT &130&2& 65& 8.6 & 3.5 &0.29& \textbf{0.21}& this study \\
\hline\\[-10pt]
\end{tabular}
$^*$RT - room temperature
\label{As}
\end{table}

Finally, we have applied the scaling law analysis for the position of lack of fusion boundary to the published experimental data for standard AlSi10Mg powders with different PSD. For this analysis, we selected studies reporting critical processing parameters that resulted in discontinuous and unstable single tracks in the experiments with AlSi10Mg powders \cite{kempen2015processing,aversa2018single,dong2020research,aboulkhair2016formation}. By considering these datasets as representing either critical energy input (for $Pe<2$), or critical enthalpy (for $Pe>2$) we determined absorptivity $A$ of these powders using equations (2) and (4), respectively. The results of this analysis are presented in Table 4.
This table compiles all data required for calculations, sourced from the referenced published studies. While both equations (2) and (4) were applied to process the data, our analysis confirms their validity only within specific Peclet number ranges: Eq. \eqref{a} for Pe $<2$ and Eq. \eqref{ab} for Pe $>2$. The values of Peclet number presented in Table 4 were calculated using the AlSi10Mg melt properties from Table 2.
One can see that applying the correct criterion for given Peclet number yields consistent absorptivity values in the range of 0.2-0.3 for all analyzed AlSi10Mg powders (these values are highlighted in bold in Table 4). It should be noted that the obtained estimates, even in cases with platform preheating (Cases 2 and 3 in Table 4), also fall within this range. 
A single outlier with $A$=0.36 was observed for Case 4 in Table 4. This discrepancy may be attributed to the approximate laser spot diameter (reported as $\approx 70\, \mu$m) in the original study \cite{dong2020research}. Considering the actual beam diameter range ($2R=40-70\, \mu$m) of the YLM-120 laser system used, the recalculated absorptivity can align well with the other results of identifications. For example, by using $2R=60\, \mu$m in Eq. \eqref{ab} for the Case 4 in Table 4 we will obtain $A=0.28$.

The results of the present study, obtained for selected combinations of critical process parameters (Cases 6 and 7 in Table 4), yield somewhat lower absorptivity  values compared to other data. This discrepancy may be attributed to two key factors. Firstly, the availability of a broader experimental dataset, which facilitates more accurate determination of the lack of fusion boundary. Secondly, the larger particle size of the powder used in the present study as compared to the other variants examined (see Table 4).

\section{Conclusions}

This paper presents the results of single-track experiments performed with  standard AlSi10Mg powder and its sieved fractions with different mean size of particles. The study yields the following key findings:

\begin{enumerate}
	\item Powder particle size significantly affects the dimensions of lack of fusion zone (and the zone of stable track formation) on the process map. The coarser AlSi10Mg powders demonstrating poorer melting behavior that shifts the melting boundary toward zones of higher input energy. At the same time, the zone of keyhole melting mode almost independent on the mean size of powder particles.
	\item The observed size effect can be attributed to the variations in powder absorptivity $A$, as confirmed by both scaling law analysis and numerical simulations. Identified values of $A$ (Tables 3, 4) should be treated as an assessment on the low energy absorptivity of the powders, since they were evaluated for relatively low level of energy input.
	\item The lack of fusion boundary is determined by the constant linear energy density for Pe $< 2$ and constant enthalpy density for Pe $> 2$. Correct predictions for the position of this boundary requires using energy-based  criterion (1) and temperature-based criterion (2), respectively, for the corresponding ranges of Peclet number. 
	\item We show the correlation between the values of absorptivity derived from the analysis of the lack of fusion boundary position and the analysis for the single-track width within the ETM model.
	\item Particle size effects on absorptivity are most pronounced at relatively low levels of laser power, at which the laser irradiation interacts primarily with the powder particles, not with the melt pool surface. This enhancement effect could potentially be leveraged in LPBF processes for highly reflective/conductive materials, for instance by using the finer powders and compensating reduced laser energy inputs with platform preheating. Presented results can be also useful for validation of advanced particle-scale models of LPBF processes.\\
\end{enumerate}

\appendix
\section{Definition for attenuation function of the heat source model and identification of its extinction coefficient}

The function $Q(\xi)$, which derivative defines the attenuation of laser power in the powder layer, was found in Ref.\cite{Gusarov2009} in the following form:
\begin{equation}
\label{cQ}
\begin{aligned}
	Q(\xi) &= \frac{R_s a}{(4R_s-3)S} 
	\Big(
	(1-R_s^2) e^{\lambda}
		\left((1-a)e^{2a\xi}+(1+a)e^{-2a\xi}
		\right)\\
		&-(3+R_s e^{2\lambda})
		(1+a-R_s(1-a))e^{2a(\xi-\lambda)}\\
		&-(3+R_s e^{2\lambda}) (1-a-R_s(1+a))e^{2a(\lambda-\xi)}
	\Big)\\
	&- \frac{3(1-R_s)(e^{\xi}-R_s e^{2\lambda-\xi})}{4R_s-3}
\end{aligned}
\end{equation}
where $R_s$ is hemispherical reflectivity of solid (particle) material, $\lambda=\chi h$ is optical thickness of the powder layer, $h$ is the thickness of powder layer, $\chi$ is the extinction coefficient, $\xi=\chi z$ is the dimensionless coordinate,  which positive direction coincides with those one of $z$ coordinate (Fig. \ref{fem1}),  
	$a=\sqrt{1-R_s}$ and 
$$S = (1-a)(1-a-R_s(1+a))e^{2a\lambda}
-(1+a)(1+a-R_s(1-a))e^{-2a\lambda}$$

The comparison between the results of numerical simulations and experimental data regarding the lack of fusion boundary position for all powder types P0-P5 is presented in Fig. \ref{femallmap}. Here, the points represent numerical calculation results where correction factors $\eta$ were adjusted to ensure close coincidence between calculated and experimental boundaries at maximum laser speed. For all other speeds, the identified $\eta$ values were used to verify model consistency with experimental data across a wide range of Peclet numbers. The absence of points in Fig. A1 indicates that the enough deep melting (with $D\approx h$) cannot be achieved in the model even with extremely high values of the correction coefficient and extinction coefficient for given set of critical process parameters established in the experiment.
The identified correction factors $\eta$ are shown in Fig. \ref{efb}a. The packing density $f$ for various investigated powders is presented in Fig. \ref{efb}b (according to Table 1). These $\eta$ and $f$ values were used to calculate the extinction coefficients $\chi$ for powders with different mean size of particles. The obtained $\chi$ values are shown in Fig. \ref{efb}c. These values were subsequently used to calculate powder absorptivity using formula \eqref{afem}.

\newpage

\setcounter{figure}{0}
 \begin{figure}[t!]
\centering
  \includegraphics[width=0.95\linewidth]{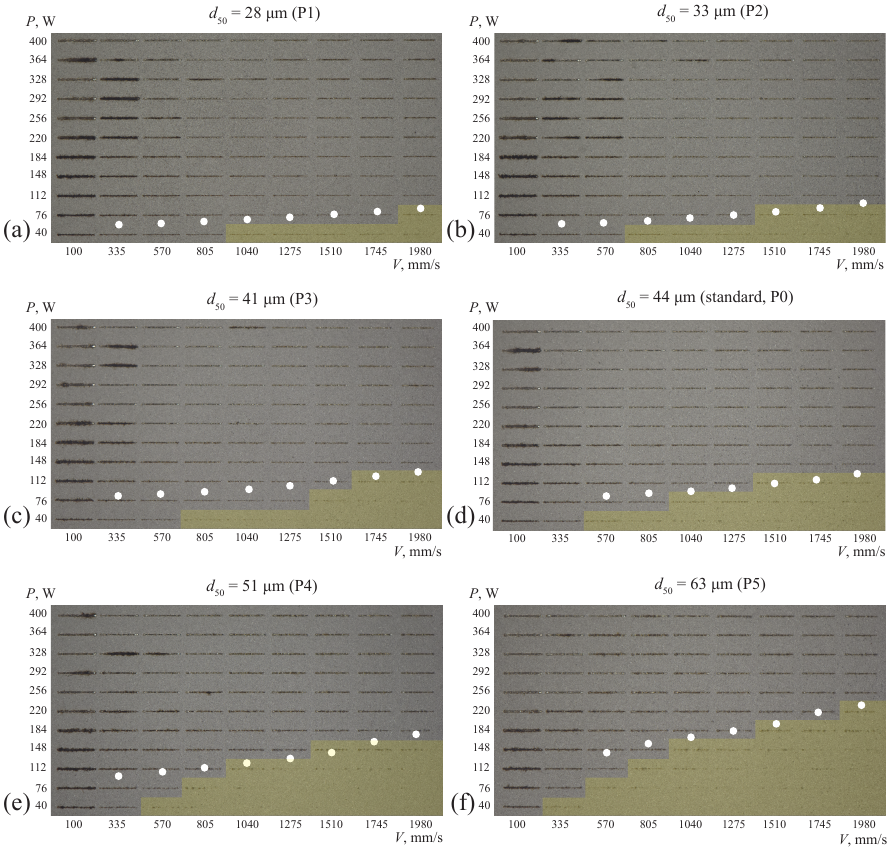}
  \caption{Comparison between the position of lack of fusion boundaries obtained in the experiments (yellow color) and predicted by the numerical simulations (white dots) for the all types of powders P0-P5 with different mean size of particles}
  \label{femallmap}
\end{figure}

\begin{figure}[h!]
\centering
  (a)\includegraphics[width=0.28\linewidth]{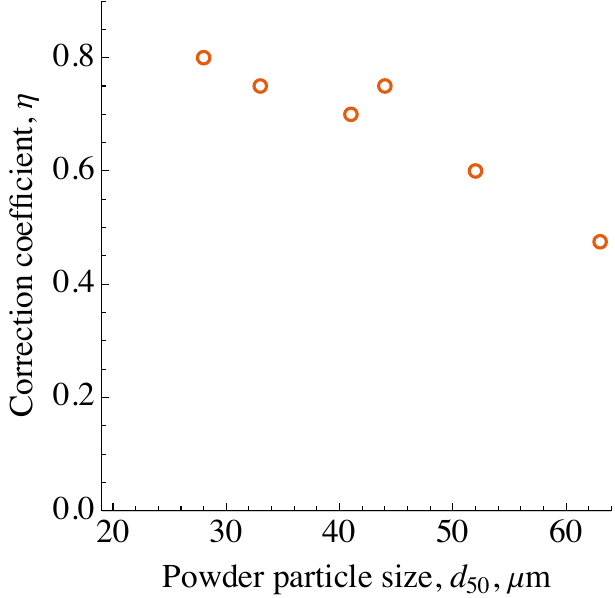}
  (b)\includegraphics[width=0.28\linewidth]{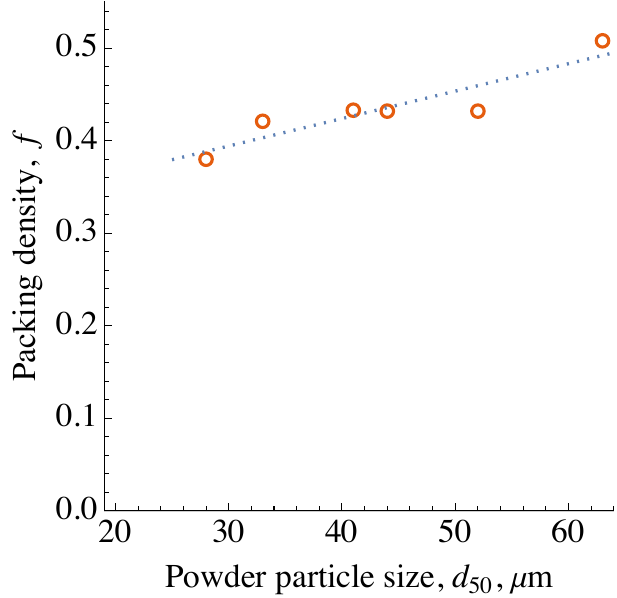}
  (c)\includegraphics[width=0.28\linewidth]{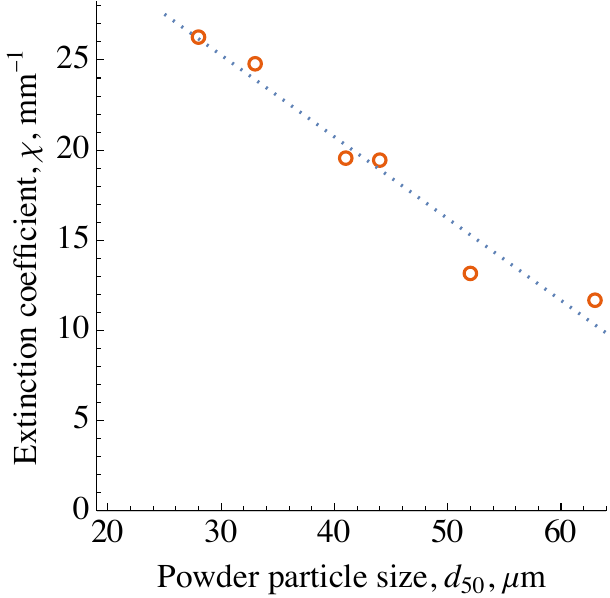}
  \caption{Dependence of identified correction coefficient (a), measured packing density of powder layer (b) and calculated extinction coefficient (c) on the mean particle size of the powders}
  \label{efb}
\end{figure}
 
 \newpage

\section*{References}
\renewcommand{\bibsection}{}
\bibliography{refs.bib}

\end{document}